\documentclass[a4paper,fleqn,usenatbib]{mnras}

\usepackage{newtxtext,newtxmath}

\usepackage[T1]{fontenc}
\usepackage{ae,aecompl}

\usepackage{graphicx}	
\usepackage{amsmath}	
\usepackage{amssymb}	
\usepackage{pifont}
\usepackage{afterpage}
\usepackage{verbatim}
\usepackage{caption}
\usepackage{subfig, float}

\title[Metals in simulated disc galaxies]{Chemical evolution of disc galaxies from cosmological simulations}

\author[M. Valentini et al.]
{Milena Valentini$^{1,2,3}$\thanks{E-mail: milena.valentini@sns.it}, 
Stefano Borgani$^{3,4,5}$\thanks{E-mail: borgani@oats.inaf.it},
Alessandro Bressan$^{2,6}$\thanks{E-mail: sbressan@sissa.it},
\newauthor
Giuseppe Murante$^{3}$,
Luca Tornatore$^{3}$, and
Pierluigi Monaco$^{3,4}$
\\ ~ \\
\footnotesize 
$^{1}$ Scuola Normale Superiore, Piazza dei Cavalieri 7, I-56126 Pisa, Italy\\
$^{2}$ SISSA - International School for Advanced Studies, via Bonomea 265, I-34136 Trieste, Italy\\
$^{3}$ INAF - Osservatorio Astronomico di Trieste, via Tiepolo 11, I-34131 Trieste, Italy\\
$^{4}$ Astronomy Unit, Department of Physics, University of Trieste, via Tiepolo 11, I-34131 Trieste, Italy\\
$^{5}$ INFN - National Institute for Nuclear Physics, Via Valerio 2, I-34127 Trieste, Italy\\
$^{6}$ INAF - Osservatorio Astronomico di Padova, Vicolo dell'Osservatorio 5, I-35122 Padova, Italy\\
}

\date{Accepted 2019 February 14.  Received 2018 December 27; in original form 2018 July 16}

\pubyear{2018}

\begin{document}
\label{firstpage}
\pagerange{\pageref{firstpage}--\pageref{lastpage}}
\maketitle

\begin{abstract}

\noindent  
We perform a suite of cosmological hydrodynamical simulations of disc galaxies, with zoomed-in 
initial conditions leading to the formation of a halo of mass $M_{\rm halo, \, DM} \simeq 2 \cdot 10^{12}$~M$_{\odot}$ 
at redshift $z=0$. 
These simulations aim at investigating the chemical evolution and the distribution of metals in a disc galaxy, 
and at quantifying the effect of {\sl{(i)}} the assumed IMF, {\sl{(ii)}} the adopted stellar yields, 
and {\sl{(iii)}} the impact of binary systems originating SNe~Ia on the process of chemical enrichment. 
We consider either a \citet{kroupa93} or a more top-heavy \citet{Kroupa2001} IMF, 
two sets of stellar yields and different values for the fraction of binary systems suitable to give rise to SNe~Ia. 
We investigate stellar ages, SN rates, stellar and gas metallicity gradients, and stellar $\alpha$-enhancement 
in simulations, and compare predictions with observations. 
We find that a \citet{kroupa93} IMF has to be preferred when modelling late-type galaxies in the local universe.
On the other hand, the comparison of stellar metallicity profiles and $\alpha$-enhancement trends with observations 
of Milky Way stars shows a better agreement when a \citet{Kroupa2001} IMF is assumed. 
Comparing the predicted SN rates and stellar $\alpha$-enhancement with observations supports a value for the 
fraction of binary systems producing SNe~Ia of $0.03$, at least for late-type galaxies and for the considered IMFs. 
Adopted stellar yields are crucial in regulating cooling and star formation, and in determining patterns 
of chemical enrichment for stars, especially for those located in the galaxy bulge.

\end{abstract}




\section{Introduction} 
\label{sec:introduction}
Metals are unique tracers of galaxy evolution and of the past history of feedback. 
Although they do not considerably contribute to the baryonic mass budget of galaxies and 
of their gaseous environments, they are a fundamental component of the galactic system. 

Metals record earlier stages of galaxy formation, since crucial processes that shape forming galaxies and determine 
their evolution leave imprints on metal distribution and relative abundance of different elements. 
Stellar feedback pollutes the interstellar medium (ISM) with heavy metals synthesised during stellar evolution, 
galactic outflows fostered by both SN (supernova) explosions and AGN (active galactic nucleus) activity spread metals 
and drive them towards the circumgalactic medium (CGM), while counterbalancing and regulating the accretion 
of pristine or metal-poor gas from the large scale environment 
\citep[see e.g. the reviews by][]{Veilleux2005, Tumlinson2017}.
Some of the enriched gas that has been driven outwards from the sites of star formation within the galaxy falls then 
back again, eventually cooling and forming subsequent generations of stars, richer in metals than previous ones. 
This feedback process involves regions of the galaxy far from the original sites where outflows originated, 
resulting in a spread and circulation of metals over galactic scales.

A number of elements contribute to determine the present day abundance and distribution of metals: the inital 
mass function (IMF), the mass-dependent lifetime function, the stellar yields, 
the fraction of stars in binary systems originating SNe~Ia, 
and the efficiency and modelling of feedback processes.
All these components are needed to build both chemodynamical models of galactic chemical evolution 
(along with the star formation history of the galaxy),
and models of chemical evolution that are included in semi-analytical models and cosmological simulations of galaxy 
formation \citep[e.g.][and references therein]{Matteucci1989, Chiappini1997, Gibson2003, Matteucci2003, Nagashima2005, tornatore2007, Borgani2008, Wiersma2009b, Yates2013, DeLucia2014, Dolag2017}.

The IMF determines the mass distribution that stars had at their birth and plays a crucial role in regulating 
the production of metals. 
Its shape sets the relative number of stars that enter the chemical evolution processes and those that do not, 
and different assumptions produce peculiar patterns of chemical enrichment. 
As for our Galaxy, different studies that have investigated the present-day mass function of stars located in 
the bulge, in the thin and thick disc agree that the IMF of MW stars, independently of their position, is 
compatible with a \citet{Kroupa2001} and \citet{ChabrierIMF2003} IMF 
\citep[see e.g.][for a recent review, and references therein]{Barbuy2018}. 

A long-standing debate concerns the universality of the IMF: observational indications suggesting that the IMF 
may vary depending on galaxy properties have been recently collected \citep[e.g.][]{Cappellari2012}. 
The possible dependence on the metallicity, density, pressure of the gas, i.e. on the physical properties of 
molecular clouds, and even on the galaxy type and environment has been investigated in different studies, 
with no general consensus \citep[e.g.][]{ChiosiBressan1998, Kroupa2013}.
The dependence of the IMF on the stellar mass in elliptical galaxies has been addressed with semi-analytical models.
\citet{Gargiulo2015} and \citet{Fontanot2017} found that a variable IMF that depends on the instantaneous 
star formation rate is suited to reproduce the observed trend of increasing $\alpha$-enhancement with larger 
galaxy stellar masses. These studies support a more top-heavy IMF in more massive systems. Similar findings are 
obtained by \citet{DeMasi2018} using a chemical evolution model. 
The impact that the variation of the IMF slopes has on star formation rates, galaxy morphology, 
chemical properties of stars and timing of chemical enrichment has been investigated also 
in simulations, where different shapes of the IMF have been considered, either assuming a global IMF 
\citep{FewCalura2014} 
or one depending on gas density and metallicity \citep{Bekki2013, GutckeSpringel2017, Barber2018}.
Besides variations of the IMF involving the low-mass end \citep{Conroy2012, LaBarbera2013}, 
also the number of massive stars predicted by the IMF has long been debated. By approximating the IMF 
with a broken power law $\phi(m) \propto m^{- \alpha}$ \citep[][for a review]{kroupa93, Kroupa2001, ChabrierIMF2003}, 
the number density of stars per mass interval can be cast as 
$\phi(m) \propto m^{-2.7}$ \citep{kroupa93} or 
$\phi(m) \propto m^{-2.3}$ \citep{Kroupa2001} for stars more massive than $1$~M$_{\odot}$, 
whether the correction for unresolved binary systems is accounted for or not 
\citep{Kroupa2002, KroupaWeidner2003}. 
The impact of binary systems has been studied by \citet{Sagar1991}, too: 
they found that the power-law slope $\alpha$ increases by $\sim 0.4$ if all stars are assumed to 
reside in binary systems. 

Stellar yields contribute to determine chemical features of gas and stars. Elements released by stars in the 
surrounding medium control radiative cooling, regulating star formation and subsequent chemical stellar feedback. 
Differences between available sets of stellar yields arise because of uncertainties on stellar nucleosynthesis, 
and due to the details of modelling stellar evolution and the structure of the star
\citep{Karakas2007, Karakas2010, Romano2010, Doherty2014a}.

Star formation and stellar deaths affect galaxies and their CGM triggering galactic outflows.  
Different stellar feedback models result in different star formation histories for simulated galaxies, 
and the way in which galactic outflows are modeled involves how metals are distributed \citep{Valentini2017}. 

Since stars retain the metals of the ISM out of which they formed, the complex interplay among the aformentioned components reflects on signatures of stars such as the 
$\alpha$-element-to-iron abundance ratio, $\alpha/Fe$. 
$\alpha$-elements (usually assumed to be O, Ne, Mg, Si, S, Ar, Ca, Ti) are metals produced as a consequence 
of helium nucleus captures during the Si burning phase just before the SNe~II core collapse. 
The $\alpha/Fe$ ratio can be used to constrain the galaxy star formation history, since the trend of $\alpha$-element 
abundance with metallicity is determined by the timescales of star formation and chemical enrichment, 
by the shape of the IMF, and by the amount of stars formed at the peak of the galaxy star formation 
history \citep{Tinsley1979}.
Stellar enhancement in $\alpha$-elements, with respect to the solar ratio, originates from typical timescales 
of chemical enrichment. Massive stars exploding as SNe~II 
pollute the ISM with $\alpha$-elements over a timescale shorter than $\sim 50$ Myr, depending on the metallicity.
On the other hand, the production of iron-peak elements is delayed with respect to $\alpha$-elements, 
since it is mainly produced by 
SNe Ia originating after the thermonuclear explosion of a white dwarf  
within a binary system. The timescale required for SN Ia explosions and subsequent chemical feedback spans 
between $\sim 50$ Myr and several Gyr.
This ratio allows us to identify different chemical evolutionary patterns for stars located in different components 
of our Galaxy, namely in the bulge, halo, thin or thick disc \citep[e.g.][]{Zoccali2006, Melendez2008}. 

A puzzling question pertains to the location of metals in and around galaxies. 
Galaxies retain some metals in their innermost regions, as heavy metals are partly locked in stars and associated to different gaseous phases of the ISM and the CGM close to the galaxy itself 
\citep[e.g.][and references therein]{Oppenheimer2012}. However, a share of the total metal budget 
is not confined to the nearest CGM and can be even lost beyond the galaxy virial radius. 
Low-density, warm CGM and IGM typically escape detections, and therefore it is difficult to probe the presence of metals several hundreds of kpc far from the galaxy centre. 
By focusing on studies that address the so-called missing metals problem 
\citep[e.g.][]{Bouche2005, Ferrara2005, Pettini2006} at redshift $z=0$, \citet{Gallazzi2008} found that 
the budget of metals locked up in stars ranges between $\sim 25 \%$ for disc-dominated galaxies 
and $\sim 40 \%$ in early-type galaxies.
\citet{Peeples2014} investigated a sample of star-forming galaxies with stellar mass in the range 
$10^9 - 10^{11.5}$~M$_{\odot}$ in the local Universe and found that galaxies retain $20 - 30 \%$ of produced metals 
in their ISM, dust and stars (neglecting metals locked up in stellar remnants). This fraction increases up to 
$50 \%$ when the CGM out to $150$ kpc is accounted for, with no significant dependence on the galaxy mass. 
Uncertainties in the adopted nucleosynthesis yields affect predictions for missing metals, since yields determine the 
total budget of heavy metals that one has to look for \citep{Peeples2014}.

Studying the distribution of heavy metals only in the innermost regions of galaxies provide us with a partial view. 
Investigating how gas flows into, within and out of galaxies allows to understand where the heavy metals eventually go. 
Cosmological hydrodynamical simulations are crucial to achieve this task, as they consistently capture the temporal and spatial complexity of gas dynamics, and account for a variety of processes, such as the chemical enrichment resulting from star formation and stellar feedback \citep[see e.g.][for a review]{Borgani2008}. 
Some light on the origin and fate of metals in different systems have been shed by analysing the chemical properties 
of the ISM, CGM, IGM (intergalactic medium) and ICM (intra-cluster medium) in simulations, and comparing them with 
observations \citep[e.g.][]{tornatore2007, Schaye2015, Oppenheimer2017, Torrey2017, Biffi2017, Vogelsberger2018}. 
Also, chemical features of stars in simulated galaxies have been investigated. For instance, \citet{Dolag2017} 
investigated the stellar metallicity as a function of the galaxy stellar mass for a sample of simulated galaxies: 
they found a stellar 
mass-metallicity relation shallower than suggested by observations, simulated galaxies with stellar mass above 
$\sim 5 \cdot 10^{10}$~M$_{\odot}$ being, on average, not as rich in iron as observed ones. Interestingly, 
\citet{Grand2016} studied how the azimuthal motion of spiral arms in simulated galaxies affect the stellar metallicity 
distribution, promoting metal-poor stars to move inward. 
Furthermore, \citet{Grand2018} investigated the metal content and $\alpha$-enhancement of stars located within 
galaxy discs, as a function of the distance from the galaxy centre and the height on the galactic plane, connecting 
chemical signatures to possible evolutionary scenarios.

The goal of this paper is to investigate the metal content of gas and stars in a suite of cosmological simulations 
of disc galaxies, taking advantage of our detailed model of chemical evolution. 
We explore the variation of the essential elements contributing to define the model of 
chemical evolution, with particular emphasis on the role played by the high-mass end shape 
of the IMF, and quantify their impact in determining metal abundance trends and evolutionary patterns. 
Guided by observations, we present here a detailed investigation focussing on the results 
at redshift $z=0$. Comparing results with observations allows us to make predictions on some of the components 
of the chemical evolution model in disc galaxies, such as the IMF and the fraction of stars in binary systems 
that are progenitors of SNe~Ia. 
The key questions that we want to address with our work are the following: 
can metallicity profiles help in supporting or discarding an IMF? 
Can the distribution of metals around simulated galaxies shed some light on the fate of metals? 
What is the impact of the different components of a chemical model on the resulting abundance pattern 
of gas and stars? 

The outline of this paper is as follows. We introduce the simulations in Section \ref{Simulations}. 
In Section \ref{Conti} we quantify the impact of two different IMFs on the predicted number of massive stars. 
In Section \ref{Results} we present and discuss our results. Section \ref{featuresOfTheGalaxy} provides an overview 
of the main features of our simulated galaxies, in Section \ref{AgeRates} we analyse stellar ages and SN rates. 
We then investigate the matal content of gas and stars, focussing on metallicity profiles (Section \ref{MetallicityProfile}) 
and stellar $\alpha$-enhancement (Section \ref{StellarContent}). 
We study the distribution of metals around galaxies in Section \ref{otherResults}, and 
the impact of adopted stellar yields in Section \ref{yields}. 
We summarise our key results and draw conclusions in Section \ref{sec:conclusions}.

\section{Numerical simulations}
\label{Simulations}

\subsection{Initial conditions and simulation setup}
\label{ICs}
 
In this work we perform a suite of cosmological hydrodynamical simulations with zoomed-in initial conditions (ICs) 
describing an isolated dark matter (DM) halo of mass $M_{\rm halo, \, DM} \simeq 2 \cdot 10^{12}$~M$_{\odot}$ at redshift $z=0$. These ICs are identified as $AqC$ and have been first introduced by \citet{Springel2008} for the DM 
component. 
The zoomed-in region that we simulate has been selected within a cosmological volume of 
$100 \, (h^{-1}$ Mpc$)^{3}$ of the DM-only parent simulation. 
We adopt a $\Lambda$CDM cosmology, with $\Omega_{\rm m}=0.25$, $\Omega_{\rm \Lambda}=0.75$, 
$\Omega_{\rm baryon}=0.04$, $\sigma_8 = 0.9$, $n_s=1$, 
and $H_{\rm 0}=100 \,h$ km s$^{-1}$ Mpc$^{-1}=73$ km s$^{-1}$ Mpc$^{-1}$. 
The Lagrangian region of the forming halo is sampled with the higher resolution DM particles. 
Each of these particles has been split into a DM plus a gas particle according to the adopted baryon fraction so as 
to have initial conditions for both DM and baryons \citep[as in][]{Scannapieco2012}. Baryons and high resolution 
DM particles define a Lagrangian volume that, by z=0, entirely contains a sphere of $\sim 3$~Mpc radius centred on the main galaxy formed (this piece of information will be exploited in Section~\ref{otherResults}). 

The simulated halo does not experience major mergers at low redshift and does not have close massive satellite 
galaxies, thus it is expected to host a disc galaxy at redshift $z=0$. 
While some of the simulated galaxies are similar to our Galaxy from the morphological point of view, our results 
should not be deemed as a model of the MW, as no specific attempts to reproduce the accretion history 
of its dynamical environment have been made. 

The simulations have been carried out with the TreePM+SPH (smoothed particle hydrodynamics) GADGET3 code, 
a non-public evolution of the GADGET2 code \citep{Springel2005}. 
In this version of the code we adopt the improved formulation of SPH introduced in \citet{beck2015}. This 
implementation accurately samples the fluid, properly describes and follows hydrodynamical instabilities, and removes 
artificial viscosity away from shock regions, as it includes a higher order kernel function, an artificial conduction term 
and a correction for the artificial viscosity. 
This new SPH formulation has been introduced in cosmological simulations adopting our sub-resolution model 
MUPPI \citep[MUlti Phase Particle Integrator,][]{muppi2010,muppi2014} according to \citet{Valentini2017}. 
MUPPI describes a multiphase ISM featuring star formation and stellar feedback, metal-dependent cooling and 
chemical enrichment, and also accounts for the presence of an ionizing cosmic background (see Section \ref{muppi}). 

In our simulations the Plummer-equivalent softening length for the computation of the gravitational force is 
$\varepsilon_{\rm Pl} = 325 \, h^{-1}$~pc, constant in 
comoving units down to $z=6$, and constant in physical units at lower redshift.
Mass resolutions for DM and gas particles are as follows: DM particles have a mass of 
$1.6 \cdot 10^6 \, h^{-1}$~M$_{\odot}$, while the initial mass of gas particles is $3.0 \cdot 10^5 \, h^{-1}$~M$_{\odot}$. 
The mass of gas particles is not constant throughout the simulation, since the initial mass can 
increase due to gas return by neighbour star particles and decrease because of star formation. 

The key features of our sub-resolution model are outlined in Section \ref{muppi}, while the simulations carried out 
for the present investigation are introduced in Section \ref{suite}.

\subsection{Stellar feedback and chemical enrichment}
\label{muppi}

Our simulations resort to the sub-resolution model MUPPI to describe processes that occur on scales 
not explicitly resolved. A thorough description of the model can be found in \citet{muppi2010, muppi2014} and 
\citet{Valentini2017, Valentini2018}: in this section we recall its most important features, while we refer the reader 
to the aformentioned papers for any further details.
Our sub-resolution model describes a multiphase ISM. Its fundamental element is the multiphase particle, that 
consists of a hot and a cold gas phases in pressure equilibrium, plus a possible stellar component. 
A gas particle enters a multiphase stage whenever its density increases above a 
density threshold ($n_{\rm thres}=0.01$ cm$^{-3}$) and its temperature drops below a 
temperature threshold ($T_{\rm thresh}=10^5$ K). 

A set of ordinary differential equations describes mass and energy flows among different components:  
hot gas condenses into a cold phase (whose temperature is fixed to $T_{c}=300$ K) due to radiative cooling, 
while some cold gas evaporates due to the destruction of molecular clouds. 
A fraction of the cold gas mass is in the molecular phase: it can be converted into stars according to a given 
efficiency that allows to compute the instantaneous star formation rate (SFR) of the multiphase particle. 
Star formation is implemented according to the stochastic model of \citet{SpringelHernquist2003}. 

Sources of energy counterbalancing radiative cooling are the energy contributed by stellar feedback (see below)
and the hydrodynamical term that accounts for shocks and heating or cooling due to gravitational compression or expansion of gas. A gas particle exits the multiphase stage whenever its density decreases below a threshold 
or after a maximum allowed time given by the dynamical time of the cold gas phase. 
A gas particle eligible to quit a multiphase stage has a probability of being kicked 
and to become a wind particle for a given time interval, during which it is decoupled from the surrounding medium. 
This probability is a parameter of our sub-resolution model (in all the simulations presented in this paper 
we assume a value of 0.03). This model relies on the assumption that galactic winds are powered by SN~II 
explosions, once the molecular cloud out of which stars formed has been destroyed. 
While being decoupled, wind particles can receive kinetick feedback energy, as described below. 

We account for stellar feedback both in thermal \citep{muppi2010} and kinetic \citep{Valentini2017} forms. 
As for thermal feedback, if a particle is multiphase, its hot gas component is heated by the energy injected 
by SN explosions within the stellar component of the particle itself, a fraction 
$f_{\rm fb, local}$ of $E_{\rm SN} = 10^{51}$~erg being deposited in its hot gas component. 
Also, SNe in neighbouring star-forming particles provide energy to both single phase and multiphase gas particles, 
this energy being supplied to the hot gas phase if the particle is multiphase. 

As for the kinetic stellar feedback, multiphase particles provide the ISM with kinetic feedback energy isotropically. 
Each star-forming particle supplies the energy $f_{\rm fb, kin} \: E_{\rm SN}$ to all wind particles within the 
smoothing length, with kernel-weighted contributions. $f_{\rm fb, kin}$ describes the kinetic stellar feedback efficiency. 
Wind particles receiving energy use it to increase their velocity along their least resistance
path, since they are kicked against their own density gradient. 
Kinetic stellar feedback is responsible for triggering galactic outflows. 
This model for kinetic stellar feedback and galactic outflows promotes the formation of disc galaxies with 
morphological, kinematic and chemical properties in keeping with observations \citep{Valentini2017, Valentini2018}. 

Star formation and evolution also produce a chemical feedback. In our model chemical evolution 
and enrichment processes are accounted for according to the model of \citet{tornatore2007}, where a detailed 
description can be found. Here we only highlight the most relevant features of the model. Star particles are 
considered to be simple stellar populations (SSPs). We evaluate the number of aging and eventually exploding stars, 
along with the amount of metals returned to the ISM, assuming an IMF and adopting stellar lifetimes and stellar yields 
(see Section \ref{suite} for details). Metals produced and released by star particles 
are distributed to neighbour gas particles within the star particles' smoothing sphere\footnote{By analogy 
	with gas particles, the mass within the sphere whose radius is the star particle smoothing length is required 
	to be constant and equal to that enclosed within the gas particles' smoothing sphere.}, 
so that subsequently generated star particles are richer in heavy elements. 
We follow in details the chemical evolution of the following elements, produced by AGB (asymptotic giant branch) 
stars, SNe~Ia and SNe~II: H, He, C, N, O, Ne, Na, Mg, Al, Si, S, Ar, Ca, Fe and Ni.
Each element independently contributes to the cooling rate, that is implemented according to \cite{wiersma2009}. 
When computing cooling rates, the effect of a spatially uniform, time-dependent ionizing cosmic background 
\citep{HaardtMadau2001} is accounted for.

\begin{table}
\centering
\begin{minipage}{\linewidth}
\caption{Relevant features of the simulation suite.
Column 1: simulation label.
Column 2: adopted IMF. 
Column 3: set of yields. 
Set A: \citet{Thielemann2003, Karakas2010, WoosleyW1995, Romano2010}.
Set B: \citet{Thielemann2003, Karakas2010, Doherty2014a, Doherty2014b, Nomoto2013}. See text for details. 
Column 4: fraction of binary systems originating SN~Ia.
Column 5: kinetic SN feedback energy efficiency.}
\renewcommand\tabcolsep{1.9mm}
\begin{tabular}{@{}lcccc@{}}
\hline
Name                              & IMF   & Set of      &      $f_{\rm bin, Ia}$     &    $f_{\rm fb, kin}$    \\
 	                              &           & yields     &           &        \\
\hline
\hline
K2s--yA--IaA--kA & \citet{Kroupa2001}   &  A              & 0.1           &   0.12\\  
  & K2s, eq (\ref{IMFonly2slopes})   &                     &             &    \\
\hline
K2s--yA--IaB--kA & \citet{Kroupa2001}   &  A                  & 0.03           &   0.12\\  
  & K2s, eq (\ref{IMFonly2slopes})   &                     &             &    \\
\hline
K3s--yA--IaA--kB & \citet{kroupa93}   &  A                  & 0.1           &   0.26\\  
  & K3s, eq (\ref{IMFslopes})   &                     &             &    \\
\hline
K3s--yA--IaB--kB & \citet{kroupa93}   &  A                  & 0.03           &   0.26\\  
  & K3s, eq (\ref{IMFslopes})   &                     &             &    \\
\hline
K2s--yB--IaB--kA & \citet{Kroupa2001}   &  B                  & 0.03           &   0.12\\  
  & K2s, eq (\ref{IMFonly2slopes})   &                     &             &    \\
\hline
K3s--yB--IaB--kB & \citet{kroupa93}   &  B                  & 0.03           &   0.26\\  
  & K3s, eq (\ref{IMFslopes})   &                     &             &    \\
\hline
\hline
\end{tabular}
\label{Simus}
\end{minipage}
\end{table}

Besides those described in Section \ref{suite} and listed in Table \ref{Simus}, all the relevant parameters of our 
sub-resolution model that we adopt for this set of simulations can be found in \citet{Valentini2018}, Table 1, 
considering the simulation labeled AqC5--fid. It is worth noting that our sub-resolution model does not have 
parameters that have been calibrated to reproduce observations of metal abundances in the ISM and CGM 
(for instance, at variance with \citet{Vogelsberger2013, Pillepich2018}, we do not adopt a wind metal loading 
factor to regulate the metallicity of the wind relative to that of the ISM from where the wind originated).

\subsection{The set of simulations}
\label{suite}

In this work we consider results of six simulations that we carried out. We list them in Table \ref{Simus}. 
These simulations have been conceived so as to investigate and quantify the effect of the adopted IMF and 
stellar yields, along with the impact of binary systems originating SN~Ia on the chemical enrichment process. 

\begin{figure}
\newcommand{\captionfonts}{\small}
\begin{minipage}{\linewidth}
\centering
\includegraphics[trim=0.2cm 0.2cm 0.0cm 0.0cm, clip, width=1.03\textwidth]{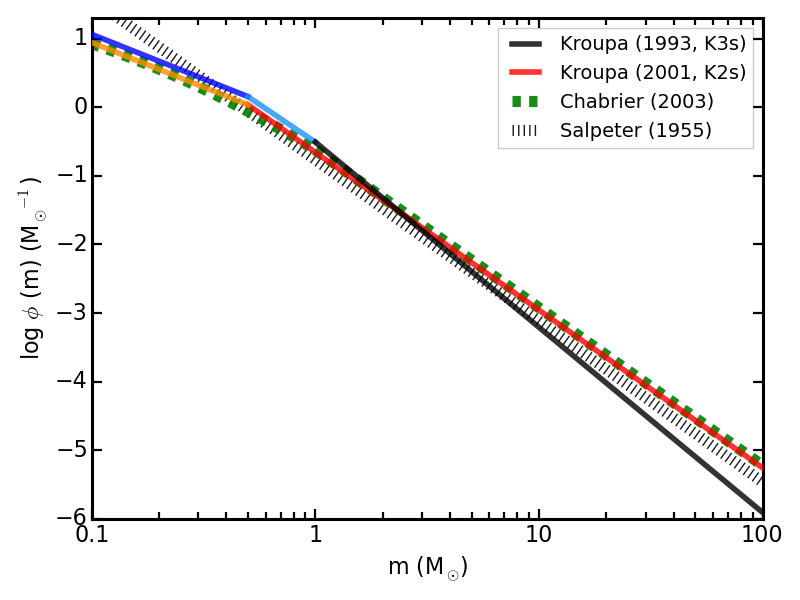} 
\end{minipage} 
\caption{Stellar number distribution per mass interval as a function of the mass 
	for the two IMFs that we are considering for the current analysis: 
	K2s is the \citet{Kroupa2001} IMF described by equation (\ref{IMFonly2slopes}), 
	while K3s is the \citet{kroupa93} IMF represented by equation (\ref{IMFslopes}). 
	For reference, we also show the \citet{ChabrierIMF2003} IMF and the \citet{Salpeter1955} IMF.}
\label{IMFpic} 
\end{figure}

We define the IMF $\phi (m)$: 
\begin{equation}
\phi (m) =  \beta m^{- \alpha}
\label{IMF}
\end{equation}
as the number of stars per unit mass interval, in the mass range $[M_{\rm inf}, M_{\rm sup}]$.
We consider two IMFs: the \citet{Kroupa2001} IMF and the \citet{kroupa93} IMF. We choose these two IMFs 
because one focus of our study is to investigate the impact of the number of massive stars on the metal 
production. We note that the \citet{Kroupa2001} IMF is almost indistinguishable by the widely adopted 
\citet{ChabrierIMF2003} IMF (see Figure \ref{IMFpic}): therefore, by considering the \citet{Kroupa2001} and 
the \citet{kroupa93} IMFs we examine the possible variation of the high-mass end shape of the IMF. 
The \citet{Kroupa2001} IMF (K2s, hereafter) is characterised by the following values for the slope $\alpha$ 
of the power law in different mass intervals: 
\begin{equation}
\begin{array}{l}
\alpha = 1.3    $\,\,\,\,\,\,\,\,\,\,$ {\text {for}} $\,\,\,\,\,\,$ 0.08 {\text { M}}_{\odot} \le m \le 0.5 {\text { M}}_{\odot}, \\
\alpha = 2.3    $\,\,\,\,\,\,\,\,\,\,$ {\text {for}} $\,\,\,\,\,\,$ 0.5 {\text { M}}_{\odot} < m \le 100 {\text { M}}_{\odot}.  \\
\label{IMFonly2slopes}
\end{array}
\end{equation}
We adopt $M_{\rm inf}=0.08$ M$_{\odot}$ and $M_{\rm sup}=100$ M$_{\odot}$ for K2s.
The IMF slope $\alpha = 2.3$ for massive stars is not corrected for unresolved stellar binaries (see below). 
For each mass interval a normalization constant $\beta$ is computed by imposing that 
$\int m \, \phi (m) \, dm = 1$ over the global mass range and continuity at the edges of subsequent mass intervals.

The second IMF that we adopt is the \citet{kroupa93} IMF (K3s, hereafter). 
This IMF is characterised by three slopes, as $\alpha$ in equation (\ref{IMF}) has the following values 
according to the mass interval:
\begin{equation}
\begin{array}{l}
\alpha = 1.3    $\,\,\,\,\,\,\,\,\,\,$ {\text {for}} $\,\,\,\,\,\,$ 0.1 {\text { M}}_{\odot} \le m \le 0.5 {\text { M}}_{\odot}, \\
\alpha = 2.2    $\,\,\,\,\,\,\,\,\,\,$ {\text {for}} $\,\,\,\,\,\,$ 0.5 {\text { M}}_{\odot} < m \le 1.0 {\text { M}}_{\odot},  \\
\alpha = 2.7    $\,\,\,\,\,\,\,\,\,\,$ {\text {for}} $\,\,\,\,\,\,$ 1.0 {\text { M}}_{\odot} < m \le 100 {\text { M}}_{\odot}.  \\
\label{IMFslopes}
\end{array}
\end{equation}
It is defined in the mass range $[0.1, 100]$ M$_{\odot}$ 
\citep[as suggested by][]{MatteucciGibson1995, Grisoni2017}. 
In this case, the slope $\alpha = 2.7$ for massive stars includes the correction for unresolved 
stellar binary systems \citep{Kroupa2002}. This correction accounts for the large fraction (more than $\sim 50 \%$) 
of stars that are located in binary systems. Surveys in our Galaxy that observe and count stars often lack the 
resolution to resolve pairs of close stars. The resulting statistics is therefore biased towards larger mass stars, 
because a fraction of them actually come from the superposition of lower mass stars. Correcting for this effect 
leads to a steepening of the slope in the high-mass range 
\citep[][see also Section \ref{sec:introduction}]{Sagar1991, Kroupa2002, KroupaWeidner2003}.

We account for different evolutionary timescales of stars with different masses by adopting the mass-dependent 
lifetimes by \citet{PadovaniMatteucci1993}. 
Stars with an initial mass larger than $M_{\rm up}$ and lower than $M_{\rm sup, \, SNII}=40$~M$_{\odot}$ are assumed to end their life exploding as core-collapse SNe.  
We consider $M_{\rm up} = 8$~M$_{\odot}$, except when the set B of yields is used (see below). In this latter case 
$M_{\rm up} = 9$~M$_{\odot}$ (see Section \ref{Conti}, too). 
Stars that are more massive than $40$ M$_{\odot}$ are assumed to implode in BHs directly, and thus do not 
contribute to further chemical enrichment or stellar feedback energy. 

A fraction of stars relative to the whole mass range is assumed to be located in binary systems that are  
progenitors of SNe~Ia. According to the model by \citet[][]{GreggioRenzini1983} and adopting the same 
formalism as in \citet{tornatore2007}, the rate of SN~Ia explosions can be cast as: 
\begin{equation}
R_{\rm SNIa} (t)=  - \frac {\text{d} m(t)}{\text{d} t} \Biggr\rvert_{\rm m_{\rm 2} \equiv \tau^{-1} (t)} \cdot 24 \, m_{\rm 2}^2 \, A \, \int^{M_{\rm B,M}}_{M_{\rm B,m}} \, \phi(m_{\rm B}) \, \frac{1}{m_{\rm B}^3} \, \text{d} m_{\rm B} \,\,\,.
\label{SN1a_rate_GR83}
\end{equation}
In equation~(\ref{SN1a_rate_GR83}), $\tau^{-1} (t)$ is the inverse of the lifetime function~$\tau(m)$, 
that describes the age at which a star of mass~$m$ dies, 
$- d m(t) / d t$ accounts for the mass of stars dying at time $t$, 
and $m_{\rm 2}$~is the mass of the secondary star in the binary system 
\citep[within the single-degenerate scenario, as in the formulation of][]{GreggioRenzini1983}. Also, 
{\sl{A}} is the fraction of stars in binary systems suitable to host SNe~Ia (see below), and 
$\phi(m)$ is the IMF. The integral in equation~(\ref{SN1a_rate_GR83}) 
is computed over the total mass of the binary system $m_{\rm B}$ producing an SN Ia, 
which varies from $M_{\rm B,m} = 3$~M$_{\odot}$ to $M_{\rm B,M} = 16$~M$_{\odot}$ 
(or from $3$ to $18$~M$_{\odot}$ when using the set B of yields)\footnote{
We note that slightly different metrics can be used when computing the number of SNe~Ia. For instance, 
\citet{Maoz2008} provides several estimates to compute the SN~Ia explosion fractions. 
Also, an interesting comparison between different SN~Ia progenitor models is presented in \citet{FewCalura2014}. 
}. 
The first SNe~Ia explode therefore in binary systems whose components have both an initial mass of $8$~M$_{\odot}$ 
(or $9$~M$_{\odot}$), according to the single-degenerate scenario. This determines the timescale required 
for SN~Ia explosions and consequent chemical enrichment, that ranges between $\sim 50$ Myr and several Gyr 
($\sim 50 \%$ of the total SNe~Ia in an SSP explode after $\sim 1$ Gyr). 
Since SNe~Ia are the main source for the production of iron-peak elements, their rate impacts directly 
on abundance trends and evolutionary patterns. In order to constraint the SN~Ia rate, we explore the variation 
of the probability of the realization of a binary system hosting a SN~Ia (while we do not consider to account for  
possible variations of stellar lifetimes or the distribution function of delay times). 
We consider two values for the fraction of binary systems giving rise to SN~Ia: either $f_{\rm bin, Ia}=0.1$ 
\citep[case A;][]{GreggioRenzini1983, MatteucciGreggio1986, tornatore2007} 
or $f_{\rm bin, Ia}=0.03$ \citep[case B; as suggested by][]{Grisoni2017}. 
This parameter is commonly referred to as {\sl{A}} (see equation~(\ref{SN1a_rate_GR83})), but we 
prefer to call it $f_{\rm bin, Ia}$ from now on (in order not to be confused with letters A and B in the simulation labels, 
see Table~\ref{Simus}). In general, it depends on the adopted IMF, 
on the systems and environments investigated \citep{MatteucciGibson1995}, 
and other values have been adopted as well \citep[e.g.][]{Spitoni2011}. 
We note that assuming the same fraction of binary systems 
when two IMFs are considered does not lead to the same number of SN~Ia explosions. 

We follow the production of different heavy metals by stars that evolve and eventually explode adopting 
stellar yields. We consider two sets of stellar yields in order to quantify the impact that they have on 
final results. 
In the first set of yields, which we refer to as set A, we assume yields tabulated by \citet{Thielemann2003} 
for SNe~Ia, while we adopt the mass- and metallicity-dependent yields by \citet{Karakas2010} for AGB stars. 
Also, we use the mass- and metallicity-dependent yields by \citet{WoosleyW1995} for SNe~II, 
combined with the prescriptions provided by \citet{Romano2010} for the Fe, Si, and S production. 
As for the second set of yields, labeled as set B, we adopt the stellar yields provided by \citet{Thielemann2003} 
for SNe~Ia and the mass- and metallicity-dependent yields by \citet{Nomoto2013} for SNe~II.
We use the mass- and metallicity-dependent yields by \citet{Karakas2010, Doherty2014a, Doherty2014b} 
for low and intermediate mass stars that undergo the AGB and super-AGB phase ($7-9$~M$_{\odot}$).
These sets of yields have been tested by state-of-the-art chemical models and well reproduce observations 
of different ion abundances in the MW \citep[][and private communications]{Romano2010, Romano2017}.
The difference between these two sets of stellar yields is quantified in Section \ref{yields}. 

The key features of our simulations are listed in Table \ref{Simus}. Besides the name and the resolution of the 
ICs, i.e. {\textsl{AqC5}}, the simulation label encodes the adopted IMF, the considered set of stellar yields, 
and two parameters, namely the value of binary systems giving rise to SNe~Ia and the kinetic stellar feedback 
efficiency, respectively.

\section{Impact of IMF on the number of massive stars}
\label{Conti}

Stellar feedback is one of the natural outcome of star formation, but it also contributes to regulate it. Therefore, 
should the number of stars that mainly contribute to stellar feedback vary as a consequence of a different IMF, 
the amount of feedback energy is directly affected, with a dramatic impact on the star formation history of the 
galaxy. The number of stars that end up their life as SNe~II deserves particular attention, as they mainly 
contribute to determine the available kinetic feedback energy and to trigger galactic outflows, expelling gas outwards. 

Figure \ref{IMFpic} shows the number density distribution per mass interval as a function of the mass 
for the two IMFs that we are considering for the current analysis (see equations (\ref{IMFonly2slopes}) and 
(\ref{IMFslopes}) for K2s and K3s IMFs, respectively).

The number of massive (m > M$_{\rm up}$) stars exploding as SNe~II in an SSP 
assuming an IMF $\phi (m) = A_{\phi} \, m^{-\alpha}$ is:
\begin{equation}
N_{\rm \ast massive} = A_{\phi}  \int_{M_{\rm up}}^{M_{\rm sup, \, SNII}} \, m^{-\alpha} \, dm \,\,\,,
\label{NumStarsIMF}
\end{equation}
$A_{\phi}$ being the normalization constant and $\alpha$ the slope of the adopted IMF 
in the considered mass range, respectively. 
In equation (\ref{NumStarsIMF}), $M_{\rm up}$ is the initial mass threshold for massive stars that explode as SNe~II 
and $M_{\rm sup}$ is the upper bound of the mass range in which the IMF is defined. 
We assume $M_{\rm sup, \, SNII} = 40$~M$_{\odot}$. 

For K2s $\phi (m) \simeq 0.22 \, m^{-2.3}$ for m > M$_{\rm up}$, while 
for K3s $\phi (m) \simeq 0.31 \, m^{-2.7}$. 
Integrating equation (\ref{NumStarsIMF}) over the mass range and assuming M$_{\rm up} = 8$~M$_{\odot}$ 
yields to the following numbers of massive stars: 
$N_{\rm \ast massive, K2s} \simeq 0.994 \cdot 10^{-2}$ for K2s, and 
$N_{\rm \ast massive, K3s} \simeq 0.497 \cdot 10^{-2}$ when K3s is adopted. 
Thus, K2s predicts roughly twice as many massive stars as K3s. 
This reflects directly in the amount of stellar feedback energy that the ISM is provided with. 
Therefore, in order to have a comparable amount of feedback energy from SN explosions, the kinetic stellar feedback 
efficiency $f_{\rm fb, kin}$ has to be doubled in simulations adopting the K3s IMF (i.e. $f_{\rm fb, kin} = 0.26$).

We note that marginal variations of the mass interval over which the integral in equation (\ref{NumStarsIMF}) is 
computed do not influence significantly the ratio of massive stars when the different IMFs are considered. 
For instance, there is not general consensus on the exact value of M$_{\rm up}$ in the range $7 - 9$~M$_{\odot}$, 
mainly because of the modelling of convection within the star. Assuming M$_{\rm up} = 9$~M$_{\odot}$ 
(instead of M$_{\rm up} = 8$~M$_{\odot}$) as the lower limit of integration in equation (\ref{NumStarsIMF}) does 
not impact significantly on the relative number of $N_{\rm \ast massive}$ nor on final results. 
We actually assume M$_{\rm up} = 9$~M$_{\odot}$ in simulations adopting the set B of stellar yields, 
as \citet{Doherty2014a, Doherty2014b} consider that stars with initial mass ranging 
between $7 - 9$~M$_{\odot}$ undergo the super-AGB phase. 



Accounting properly for the number of SNe~II and recalibrating the feedback efficiency accordingly allows us 
to have a comparable star formation history for simulations K2s--yA--IaA--kA and 
K3s--yA--IaA--kB (see Figure \ref{sfr} and Section~\ref{featuresOfTheGalaxy}). 
We emphasise that this is the reason why we decide to contrast results from simulations adopting 
the K2s IMF and $f_{\rm fb, kin} = 0.12$ ({\sl{kA}} in simulation names), and the ones with the K3s IMF 
and $f_{\rm fb, kin} = 0.26$ ({\sl{kB}} in simulation labels) when analysing the effect of the varying IMF. 
The total stellar feedback energy is still reduced when the K3s IMF is assumed. Nonetheless, 
we are interested primarily in the kinetic stellar feedback, as it is responsible for the triggering of galactic outflows, whose role is key in controlling gas accretion, gas ejection, and star formation \citep[see also][]{Valentini2017}.

\begin{figure*}
\newcommand{\captionfonts}{\small}
\begin{minipage}{\linewidth}
\centering
\vspace{-2.ex}
\includegraphics[trim=0.cm 0.cm 0.cm 0.cm, clip, angle=270, width=1.\textwidth]{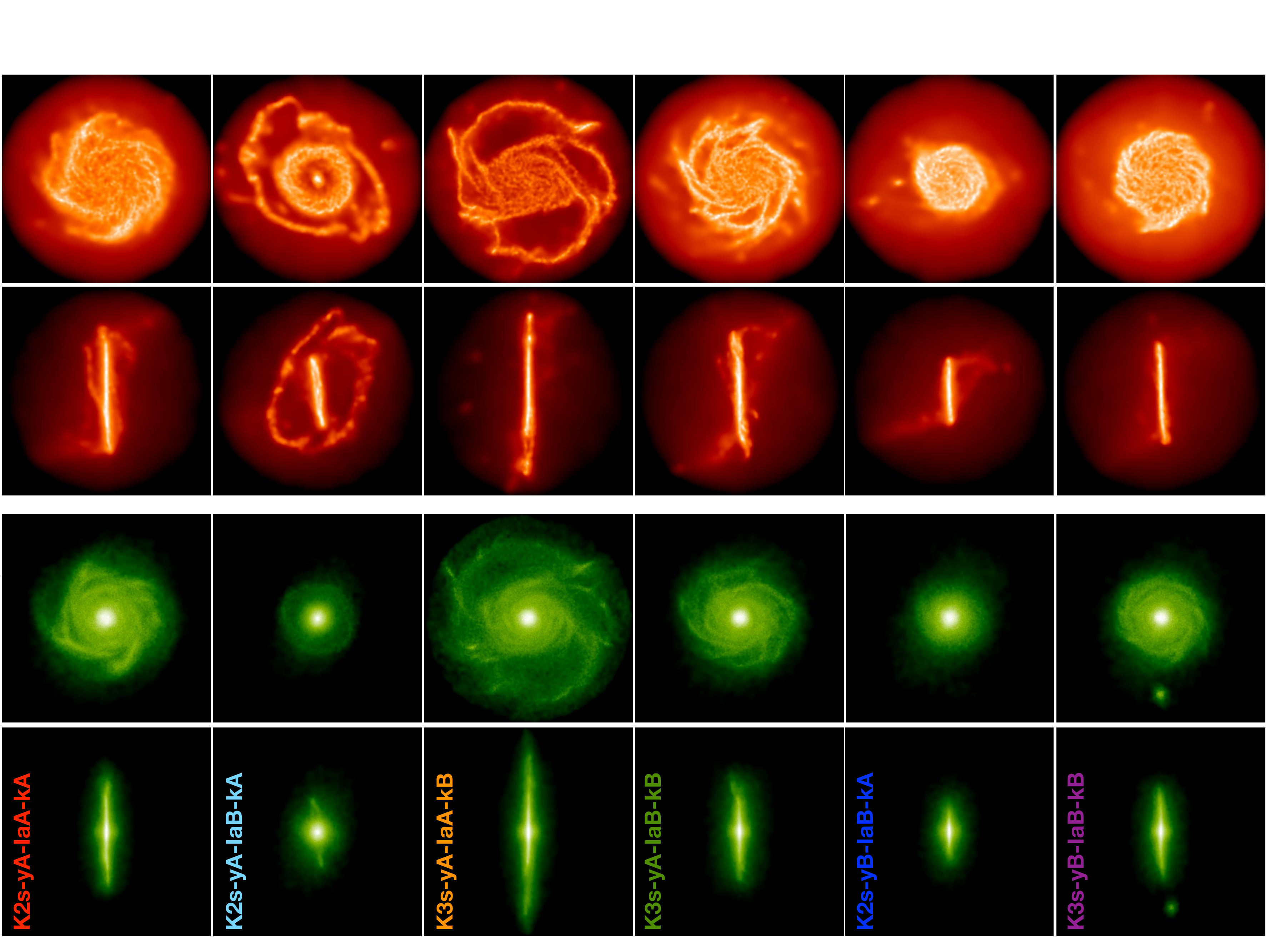} 
\end{minipage} 
\caption{Projected stellar (first and second columns) and gas (third and forth columns) density maps for the set 
	of simulated galaxies listed in Table \ref{Simus}, at redshift $z = 0$. 
	Each row shows a galaxy, whose name is indicated in the first column panel. First and third columns 
	show edge-on galaxies, second and forth columns depict face-on maps. The size of each box is 48 kpc a side.}
\label{StellarDensityMaps} 
\end{figure*}

At variance with our aformentioned recalibration, \citet{GutckeSpringel2017} did not tune the stellar feedback efficiency accounting for the different number of stars produced by the adopted metallicity-dependent IMF, with a consequent overproduction of stellar mass resulting from a modified star formation history.

\section{Results}
\label{Results}

In this section we present our results. In Section \ref{featuresOfTheGalaxy} we show the main properties 
of the simulated galaxies. In Section \ref{AgeRates} we analyse stellar ages and SN rates, while 
in Section \ref{MetallicityProfile} we show radial abundance profiles for gas and stars in our set of galaxies. 
In Sections \ref{StellarContent} and \ref{otherResults} we investigate the $\alpha$-enhancement of stars and 
where metals are located within and around our galaxies. 
In Section \ref{yields} we investigate how stellar yields impact on final results. 

Throughout this paper, when we mention metallicity we refer to the abundance by number of a given element or 
ratio between two elements. When comparing element abundance to that of Sun, we adopt values 
for the present-day Sun's abundance in the element $X$ 
(i.e. log$_{10} \varepsilon_{\rm X} = 12 \, +$ log$_{10} \bigl({N_{\rm X}}/{N_{\rm H}} \bigr)$, where 
$N_{\rm X}$ and $N_{\rm H}$ are number densities of the element $X$ and of hydrogen, respectively) 
according to \citet[][as for iron]{Caffau2011} and to \citet[][for all the other elements]{Asplund2009}.

\subsection{Main properties of the simulated galaxies}
\label{featuresOfTheGalaxy}

We introduce the main physical properties of the galaxies resulting from our cosmological simulations. 
Figure \ref{StellarDensityMaps} shows projected stellar (first and second columns) and 
gas (third and forth columns) density maps for each galaxy. 
Both edge-on (first and third columns) and face-on (second and forth columns) views are presented. 
Galaxies have been rotated so that the z-axis of their reference system is aligned with the angular momentum
of star particles and cold and multiphase gas particles located within $8$~kpc from the minimum 
of the gravitational potential. 
The centre of the galaxy, where the origin of the reference system is set, is assumed to be the centre of mass 
of the aformentioned particles.
Here and in the following, we consider for our analysis star and gas particles that are located within the 
galactic radius\footnote{We define here the galactic 
	radius as one tenth of the virial radius, i.e. $R_{\rm gal}= 0.1 R_{\rm vir}$. We choose 
  	the radius $R_{\rm gal}$ so as to identify and select the region of the 
  	computational domain that is dominated by the central galaxy.
  	Moreover, we consider virial quantities as those computed in a sphere that is centred 
	on the minimum of the gravitational potential of the halo and that encloses 
	an overdensity of 200 times the {\sl critical} density at present time.}, unless otherwise stated. The 
galactic radius of these galaxies is $R_{\rm gal} \sim 24$ kpc.
	
Density maps in Figure \ref{StellarDensityMaps} show that all the galaxies 
but K2s--yA--IaB--kA and K2s--yB--IaB--kA have a 
limited bulge and a dominant disc component, although the morphology and the extent of the disc varies from galaxy 
to galaxy. In particular, some galaxies exhibit an irregular distribution of gas around and above the galactic plane, 
this suggesting ongoing gas accretion (see also Figure \ref{mah}). A clear spiral pattern is evident in the majority of 
the discs.

\begin{figure}
\newcommand{\captionfonts}{\small}
\begin{minipage}{\linewidth}
\centering
\includegraphics[trim=0.3cm 0.cm 0.2cm 0.2cm, clip, width=1.\textwidth]{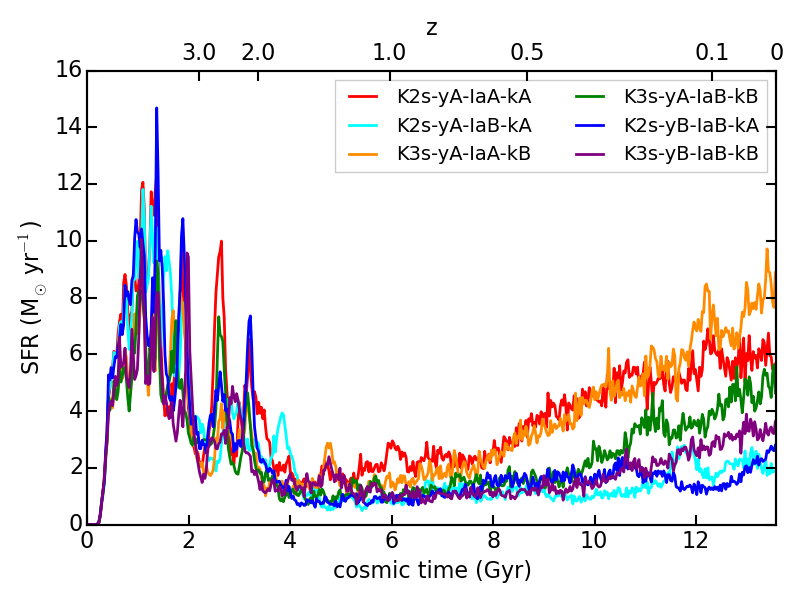} 
\end{minipage} 
\caption{Star formation histories for the six simulated galaxies listed in Table \ref{Simus}. 
	Red curve shows K2s--yA--IaA--kA, cyan and orange lines depict K2s--yA--IaB--kA and
	K3s--yA--IaA--kB, respectively. The green curve refers to K3s--yA--IaB--kB, while 
	blue and purple lines show K2s--yB--IaB--kA and K3s--yB--IaB--kB, respectively
	(as shown in the legend).}
\label{sfr} 
\end{figure}

\begin{figure}
\newcommand{\captionfonts}{\small}
\begin{minipage}{\linewidth}
\centering
\includegraphics[trim=0.3cm 0.cm 0.1cm 0.cm, clip, width=1.\textwidth]{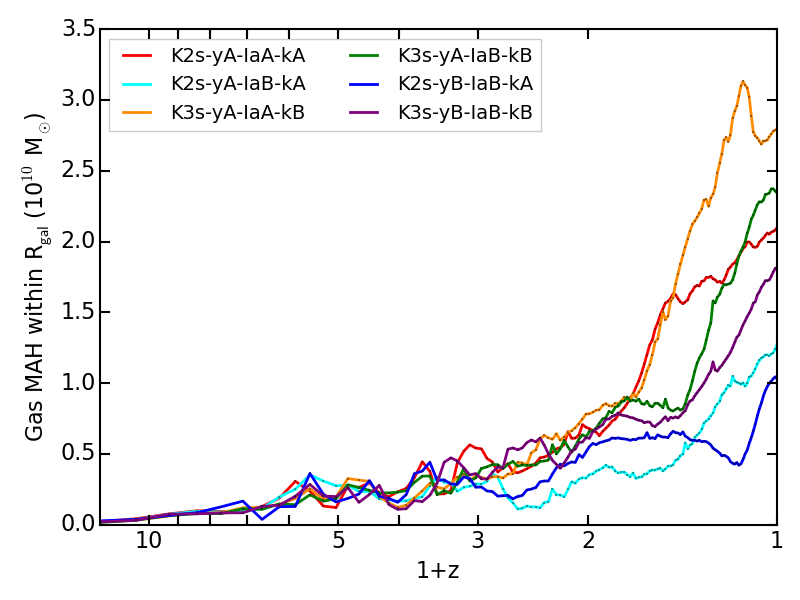} 
\end{minipage}  
\caption{Mass accretion history (MAH) of gas within the galactic radius $R_{\rm gal}$ of galaxies. We show the redshift evolution of the mass of gas that is within the $R_{\rm gal}$ of the main progenitor of our 
simulated galaxies. Colours as in Figure \ref{sfr}.}
\label{mah} 
\end{figure}

\begin{figure}
\newcommand{\captionfonts}{\small}
\begin{minipage}{\linewidth}
\centering
\includegraphics[trim=0.3cm 0.cm 0.1cm 0.cm, clip, width=1.\textwidth]{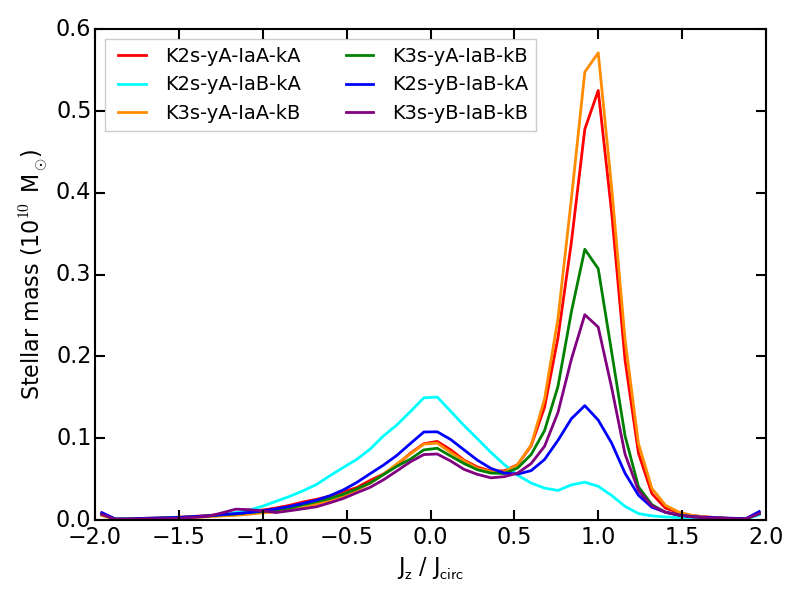} 
\end{minipage} 
\caption{Stellar mass as a function of the circularity of stellar orbits at $z=0$ for the set of galaxies. The height 
	of the peaks at $J_{\rm z}/J_{\rm circ}=0$ and at $J_{\rm z}/J_{\rm circ}=1$ shows the relative contribution 
	to the total stellar mass of the bulge and the disc, respectively. Bulge-over-total mass ratios are as follows: 
	$0.33$ (red), $0.97$ (cyan), $0.28$ (orange), $0.41$ (green), $0.64$ (blue), and $0.45$ (purple).}
\label{jcirc} 
\end{figure}

In Figure \ref{sfr} we show the evolution of the SFR of the six galaxies. 
A high-redshift ($z \gtrsim 3$) star formation burst characterises the star formation history of all the galaxies, 
in a remarkably similar way, building up the bulk of the stellar mass in the bulge of each galaxy. 
Star formation occurring at lower redshift, in a more continuous way, is then responsible of the (possible) formation 
of the disc. 
General trends are as follows: the lower the fraction of binary systems originating SNe~Ia, the lower the SFR 
below $z \sim 1$. Also, the set B of stellar yields is responsible for the reduced gas cooling and consequent lower SFR. 
This is particularly evident when contrasting the SFR of K3s--yA--IaB--kB and K3s--yB--IaB--kB.
The reason stems from the lower amount of iron that is produced when reducing the number of SNe~Ia, 
the iron being one of the main coolants \citep[along with oxygen, for solar abundances,][]{wiersma2009}. 
Also, the stellar yields labeled as B predict a lower synthesis of Fe for all the considered metallicities and a lower 
production of O for solar metallicity (see Sections \ref{StellarContent} and \ref{yields}). 
Effects on the star formation are evident, as a consequence of the element-by-element metal cooling in our 
simulations.

A complex interplay between different processes determines and regulates the star formation histories of 
the galaxies in Figure~\ref{sfr}: the reservoir of gas available for star formation, the cooling process, and 
stellar feedback. Figure~\ref{mah} shows the mass accretion history of gas within the galactic radius 
$R_{\rm gal}$ for the main progenitor of each galaxy: 
it quantifies the mass of gas within $R_{\rm gal}$, and how inflows and outflows 
impact on the amount of gas within $R_{\rm gal}$ across cosmic time. It can be related 
to the evolution of the SFR. However, there is not a direct correspondance between the amount 
of accreted gas and the SFR. The reason is that gas accreted becomes available for star formation as soon as it 
approaches the innermost regions of the forming galaxy and cools. Besides on density, gas cooling depends on 
metallicity, that primarly shapes the cooling function. Also, massive stars trigger galactic outflows as they explode 
as SNe II: outflows can hamper gas that is accreted within the virial radius from flowing inwards and feed star 
formation. A more top-heavy IMF (i.e. K2s) would further affect this scenario if the feedback efficiency were not 
calibrated (see Section \ref{Conti}). We note that minor variations between low-redshift 
star formation histories and gas mass accretion histories of galaxies with different IMFs are present, 
despite the feedback efficiency calibration: this is a further evidence of how different elements 
enter the process of galaxy formation in a non-linear fashion. Also, the formation of the stellar disc is highly 
sensitive to the timing of gas accretion \citep[see also][]{Valentini2017}.


\begin{table*}
\centering
\begin{minipage}{\linewidth}
\caption{Success of galaxy models in matching observational constraints considered 
throughout Section~\ref{Results}. Each row refers to a galaxy model of our simulation suite. 
Columns are the observational tests.}
\renewcommand\tabcolsep{4.mm}
\newcommand{\cmark}{\ding{51}}
\newcommand{\xmark}{\ding{55}}
\begin{tabular}{@{}lcccccc@{}}
\hline
Name                              & Dominant stellar         & SN Ia    &    SN II     &  Gas    &    Stellar      &   $\alpha$-enhancement    \\
 	                              & disc component          &   rates   &    rates   &  metallicity gradient  &   metallicity gradient  &   pattern   \\
\hline
\hline
K2s--yA--IaA--kA  & \cmark & \cmark & \xmark & \xmark & \cmark & \xmark \\
\hline
K2s--yA--IaB--kA & \xmark & \cmark & \cmark & \xmark & \cmark & \cmark \\ 
\hline
K3s--yA--IaA--kB & \cmark & \cmark & \xmark & \cmark & \cmark & \xmark \\
\hline
K3s--yA--IaB--kB & \cmark & \cmark & \cmark & \cmark & \cmark & \cmark \\
\hline
K2s--yB--IaB--kA & \xmark & \cmark & \cmark & \cmark & \cmark & \cmark \\
\hline
K3s--yB--IaB--kB & \cmark & \cmark & \cmark & \cmark & \xmark & \cmark \\
\hline
\hline
\end{tabular}
\label{Success}
\end{minipage}
\end{table*}


Figure \ref{jcirc} presents a kinematic decomposition of simulated galaxies. We examine the circularity of stellar orbits 
in order to distinguish between the rotationally and dispersion supported components of each galaxy, thus quantifying 
the prominence of the disc with respect to the bulge. The circularity of stellar orbits is defined as the ratio of specific 
angular momenta $J_{\rm z}/J_{\rm circ}$, where $J_{\rm z}$ is the specific angular momentum in the direction 
perpendicular to the disc, and $J_{\rm circ}$ is the angular momentum of a reference circular orbit at the 
considered distance from the galaxy centre \citep{Scannapieco2009}. Stars located in the disc contribute to the 
peak where $J_{\rm z}/J_{\rm circ}=1$, while stars in the bulge are characterised by $J_{\rm z}/J_{\rm circ} \sim 0$. 
Galaxies K3s--yA--IaA--kB and K2s--yA--IaA--kA exhibit the most pronounced disc, 
as a consequence of the higher low-redshift SFR (see Figure \ref{sfr}). An extended disc where the bulk of 
the stellar mass is located characterises galaxies K3s--yA--IaB--kB and K3s--yB--IaB--kB, 
while K2s--yB--IaB--kA has a smaller stellar disc component. 
Finally, K2s--yA--IaB--kA is an irregular galaxy with a dominant spheroidal component 
(as shown in Figure \ref{StellarDensityMaps}). 
Table~\ref{Success} summarises the morphological features of different galaxy models and 
whether galaxy properties meet or not observational constraints that we consider during our analysis 
throughout Section~\ref{Results}. 
The prominence of the stellar disc can be easily explained by analysing the SFR below $z \sim 1$ in 
Figure \ref{sfr}. A higher low-redshift SFR translates directly in a more extended stellar disc component. 
Stellar masses of galaxies and bulge-over-total (B/T) mass ratios are as follows: 
$3.64 \cdot 10^{10}$~M$_{\odot}$ and 0.33 (K2s--yA--IaA--kA), 
$2.02 \cdot 10^{10}$~M$_{\odot}$ and 0.97 (K2s--yA--IaB--kA), 
$3.85 \cdot 10^{10}$~M$_{\odot}$ and 0.28 (K3s--yA--IaA--kB), 
$2.68 \cdot 10^{10}$~M$_{\odot}$ and 0.41 (K3s--yA--IaB--kB), 
$2.06 \cdot 10^{10}$~M$_{\odot}$ and 0.64 (K2s--yB--IaB--kA), 
$2.24 \cdot 10^{10}$~M$_{\odot}$ and 0.45 (K3s--yB--IaB--kB).

We note that our estimate for 
	the B/T ratio could also include satellites, stellar streams, and contribution from bars within $R_{\rm gal}$. 
	Therefore, the B/T values that we quote should not be directly compared with observational photometric 
	estimates, as the photometric determination for the value of B/T has been shown to be lower than 
	the corresponding kinematic estimate \citep{Scannapieco2010}.

\begin{figure}
\newcommand{\captionfonts}{\small}
\begin{minipage}{\linewidth}
\centering
\includegraphics[trim=0.1cm 0.0cm 0.2cm 0.0cm, clip, width=1.\textwidth]{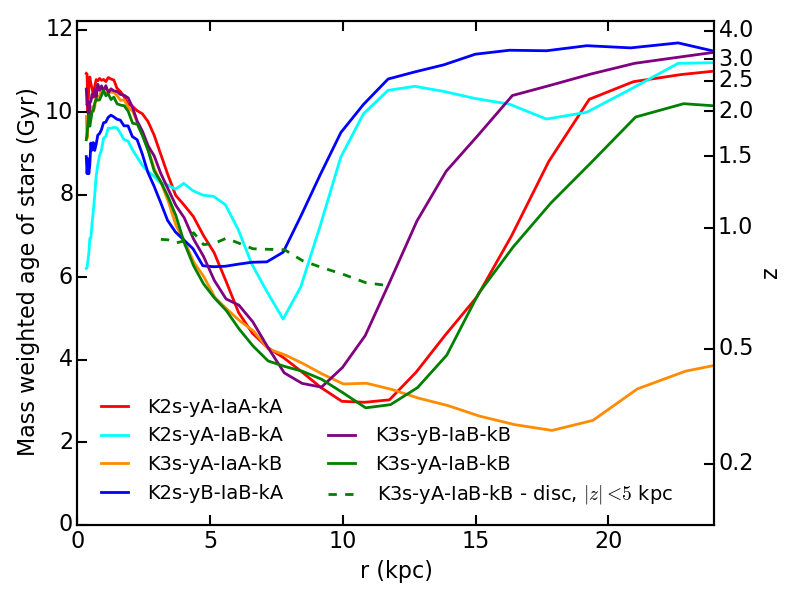} 
\end{minipage} 
\caption{Radial profiles of the mass weighted age of stars within the galactic radii of the six simulations. For 
	simulation K3s--yA--IaB--kB (green) we also show the mass weighted age of stars in the disc (dashed line).}
\label{StellarAgeProfiles} 
\end{figure}

\subsection{Stellar ages and SN rates}
\label{AgeRates}

We continue our analysis by investigating stellar ages and formation epochs for different components of our galaxies. 
Figure \ref{StellarAgeProfiles} shows radial profiles of the mass weighted age of stars within the galactic radius for 
the six simulations. The corresponding redshift at which stars form is displayed, too. 
Stars in the innermost regions of galaxies have, on average, an age ranging between $9.5$ and $11$~Gyr, depending 
on the simulation. The mass weighted mean age of stars progressively decreases as the distance from the galaxy 
centre increases and a minimum is reached in correspondence with the extent of the stellar disc of each galaxy. 
Moving then outwards, the mean stellar age of the stars progressively increases. 
When computing the age of stars located at increasing distance from the galaxy centre, we consider stars within 
spherical shells with a given radial thickness. Therefore, albeit the identification of the innermost and outermost old 
stars with bulge and halo stars, respectively, is straightforward, stars that make the radial profile decline cannot be 
associated directly to the disc component.

\begin{figure*}
\newcommand{\captionfonts}{\small}
\begin{minipage}{\linewidth}
\centering
\includegraphics[trim=0.4cm 0.0cm 1.15cm 0.0cm, clip, width=0.497\textwidth]{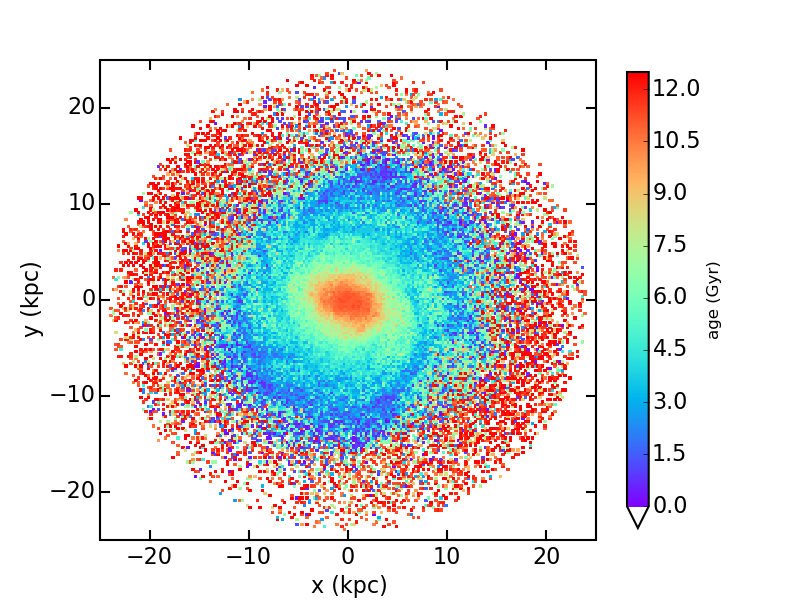} 
\includegraphics[trim=0.4cm 0.0cm 1.15cm 0.0cm, clip, width=0.497\textwidth]{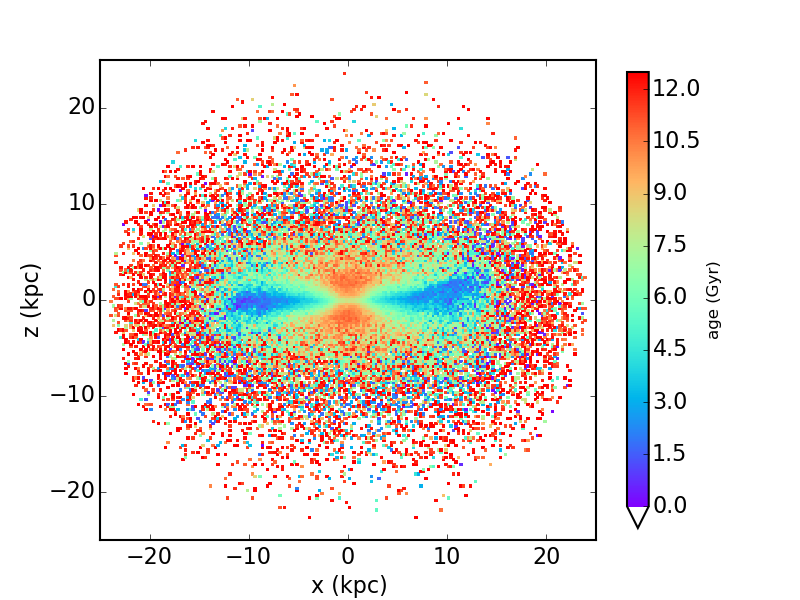} 
\end{minipage} 
\caption{Face-on (left-hand panel) and edge-on (right-hand panel) binned distributions of all the star particles located 
within the galactic radius for the K3s--yA--IaB--kB (green) simulation. Plots are shown at redshift $z = 0$; 
the colour encodes the mean age of the star particles in the bin.}
\label{ageDensityMaps} 
\end{figure*}

\begin{figure*}
\newcommand{\captionfonts}{\small}
\begin{minipage}{\linewidth}
\centering
\includegraphics[trim=0.1cm 0.0cm 0.2cm 0.0cm, clip, width=1.\textwidth]{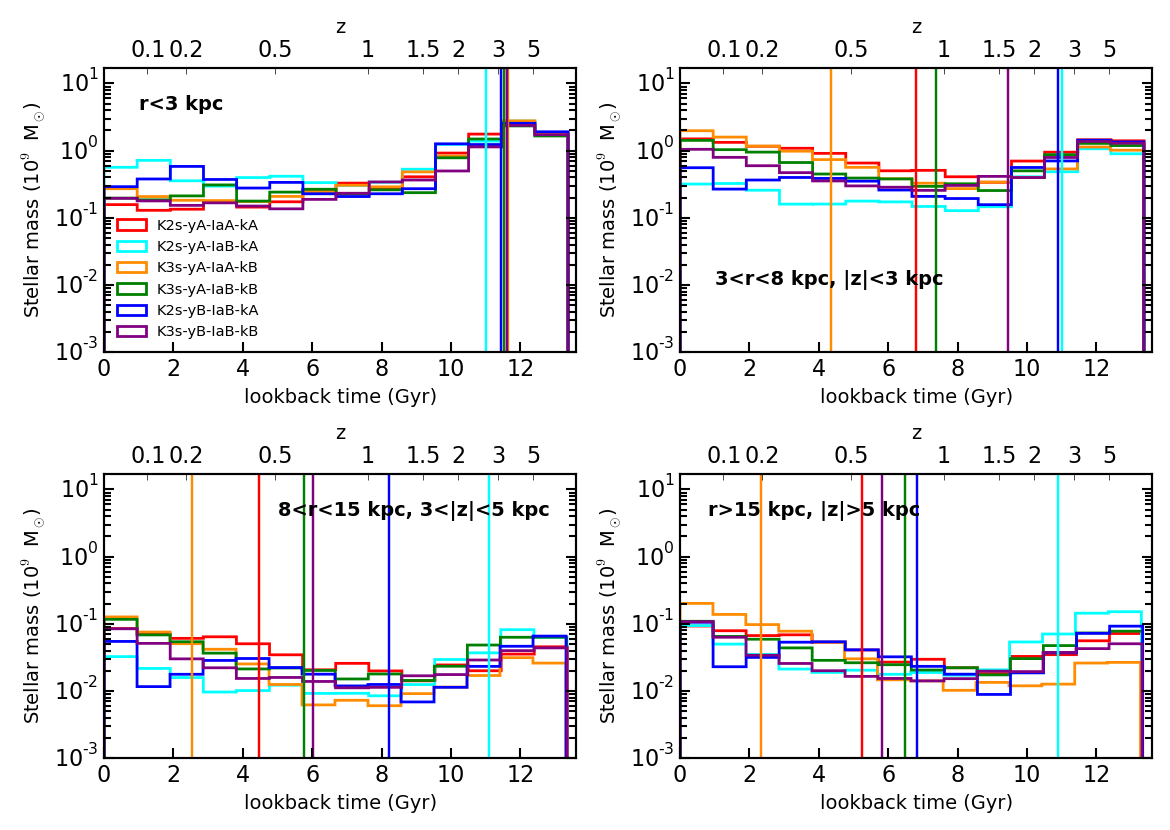} 
\end{minipage} 
\caption{Distribution of stellar mass per bin of lookback time for all the simulations. The four panels consider 
	different regions of the volume $r < R_{\rm gal}$, as reported in each panel, and roughly identify the following 
	components of the galaxy: bulge (top left), inner disc (top right), outer disc (bottom left), and halo (bottom right). 
	Vertical lines show the median of each distribution, and highlight the formation epoch of different component.}
\label{StellarAgeHisto} 
\end{figure*}

\begin{figure*}
\newcommand{\captionfonts}{\small}
\begin{minipage}{\linewidth}
\centering
\includegraphics[trim=0.4cm 0.0cm 0.15cm 0.0cm, clip, width=0.497\textwidth]{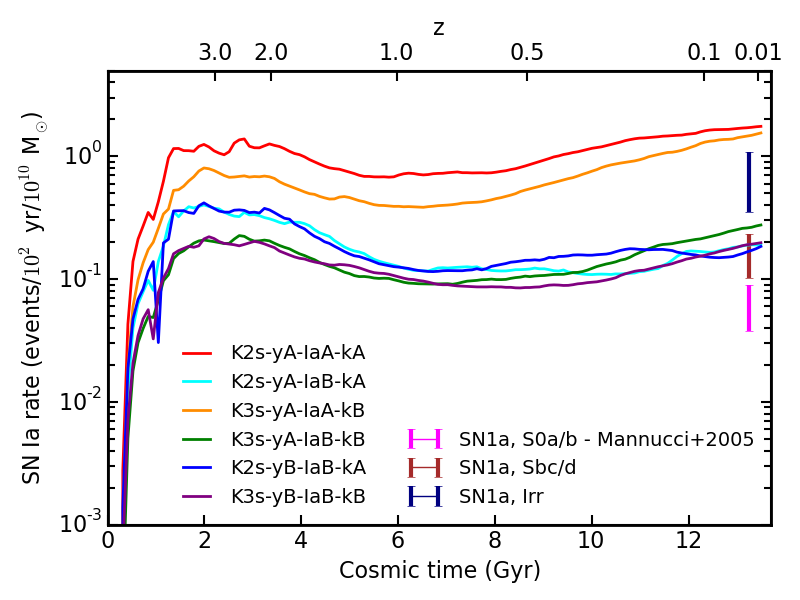} 
\includegraphics[trim=0.4cm 0.0cm 0.15cm 0.0cm, clip, width=0.497\textwidth]{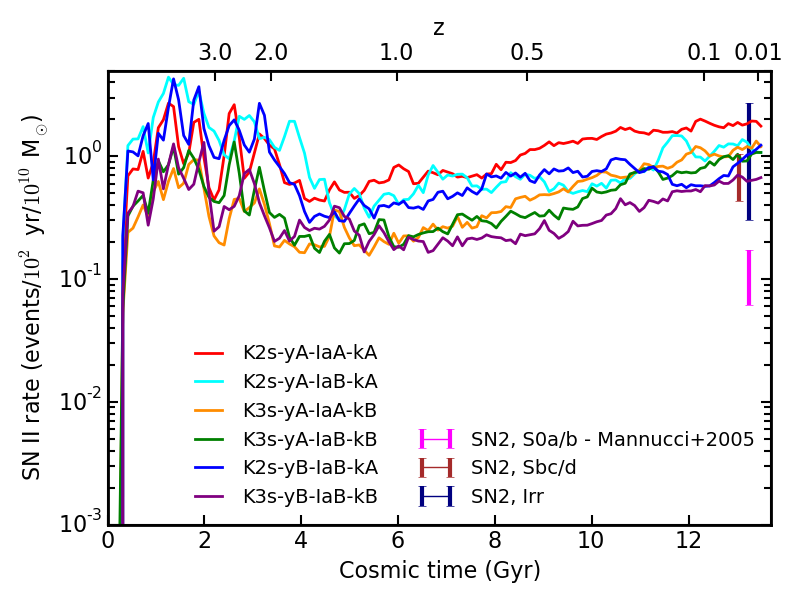} 
\end{minipage}
\caption{Evolution of SN~Ia (left-hand side panel) and SN~II (right-hand side panel) rates for the simulated galaxies. 
	Rates are expressed as events per century and per $10^{10}$~M$_{\odot}$. Our predictions are 
	compared with observations by \citet{Mannucci2005} for local universe galaxies with different morphological type, 
	as highlighted in the legend.} 
\label{SNrate} 
\end{figure*}

For this reason, in order to further investigate stellar ages for stars in the disc, we focus on simulation 
K3s--yA--IaB--kB and consider a region limited as follows: $3 < r < 13$ kpc and $|z| < 5$ kpc, $r$ and $z$ being 
the radial distance from the galaxy centre and the galactic latitude, respectively. 
We show the mass weighted age of stars in the aformentioned volume that can be deemed as the disc 
(green dashed line) in Figure~\ref{StellarAgeProfiles}, too. Stars are characterised by a negative age radial gradient, 
and this supports the inside-out growth for the stellar disc. This scenario, where stellar discs progressively increase 
their size as they evolve and grow more massive, is corroborated by a variety of observations 
\citep[e.g][]{Barden2005, MunozMateos2011, Patel2013, GonzalezDelgado2014}, 
and by models of chemical evolution \citep[e.g.][]{Chiappini2001}, by semi-analytical models 
\citep[e.g.][]{Dutton2011a}, and simulations \citep[e.g.][]{Pichon2011, Aumer2014}. 
We chose the K3s--yA--IaB--kB as a reference model to further investigate stellar ages in the galaxy disc. 
Comparable conclusions can be drawn also for other galaxy models with extended stellar discs: however, 
we preferred to overplot only one further radial profile in Figure~\ref{StellarAgeProfiles} for the sake of clarity. 
Also, restricted regions where inspecting the age of stars located in the disc have to take into account the 
galaxy morphology, so that it is not trivial to define only one region for all the galaxy models 
where studying stars in the disc. Note that the stellar age profile of stars in the disc of K3s--yA--IaB--kB is flatter than 
the profile characterising all the stars in that considered range of distance from the galaxy centre, 
younger stars at higher latitudes on the galactic plane being not included. 

Inspecting the stellar age of K3s--yA--IaB--kB further, we show the face-on and edge-on distribution 
of all the star particles within the galactic radius $R_{\rm gal}$ at redshift $z = 0$ in Figure \ref{ageDensityMaps}. 
The colour encodes the mean age of the star particles in each spatial bin: we can 
appreciate how the stellar age changes as a function of the distance from the galaxy centre across the 
galactic plane (left-hand panel of Figure \ref{ageDensityMaps}) and as a function of the galactic latitude 
(right-hand panel). 
The youngest stars trace the spiral pattern of the galaxy, and pinpoint spiral arms as sites of recent star formation. 
The thickness of the stellar disc progressively increases moving outwards. The halo and the bulge are made up of the 
oldest stars in galaxy. 

When analysing stellar ages at $z=0$ in Figures \ref{StellarAgeProfiles} and \ref{ageDensityMaps}, we analyse 
mean quantities, thus it is not trivial to deduce straightforwardly the formation scenario. 
If we consider the distribution in Figure \ref{ageDensityMaps} at higher-redshift simulation outputs 
(i.e. by evolving it backward in time), we find that high-z star formation occurs first in the innermost regions 
of the forming galaxy and soon after it is widespread in the galaxy, the oldest ($\sim 12$~Gyr) stars spreading 
across the halo as well as in inner regions \citep[see][for a similar scenario in elliptical galaxies]{ClemensBressan2009}. 
As time passes, star formation involves internal regions alone, as they are replenished with gas that is accreted from 
the large scale environment, and takes place especially in the forming bulge and at relatively larger distances from the 
centre with few episodes of star formation. 
Then, at later epochs ($z \lesssim 1.5$), star formation in the disc occurs and the disc formation proceeds inside-out.

The presence of an old ($\sim 12$~Gyr) stellar component throughout the galaxy can be seen in 
Figure \ref{StellarAgeHisto}, where the distribution of stellar mass per bin of lookback time is shown. 
The galaxy stellar content has been sliced into four different regions (within the galactic radius, $r < R_{\rm gal}$), 
as reported in each panel. 
The regions can be considered as representative of the following components of the galaxy: bulge (top left), 
inner disc (top right), outer disc (bottom left), and halo (bottom right). Vertical lines show the median of each 
distribution, and highlight the formation epoch of different components. 

We then analyse SN rates in our set of simulated galaxies.
Figure \ref{SNrate} shows the redshift evolution of SN~Ia (left-hand side panel) and SN~II (right-hand side panel) rates, where 
rates are expressed as events per century and per $10^{10}$~M$_{\odot}$. We compute rates as SN events between 
two simulation outputs over the time between the outputs. 
We compare predictions from our simulations with observations by \citet{Mannucci2005}. These authors compute the rate of SNe for galaxies belonging to different morphological classes: here we consider SN~Ia and SN~II rates for 
galaxies classifies as irregular, S0a/b and Sbc/d in the observational sample. Error bars represent the $1-\sigma$ 
error (including uncertainties on SN counts, and host galaxy magnitude and inclination), but do not account for 
uncertainties on the estimate of the mass for galaxies in the observational sample, that is quoted to affect 
measurements by $\sim 40\%$. 

By comparing Figures \ref{SNrate} and \ref{sfr}, we can appreciate how SN~II rates trace the evolution of the SFR. 
We find that SN~II rates of our galaxies are in keeping with observations from late-type Sbc/d galaxies, 
the galaxy model K2S--yA--IaA--kA slightly exceeding them. 
Also, we find a good agreement between SN~Ia rates of simulations adopting the lower fraction of binary systems 
that host SNIa and observations from Sbc/d galaxies. 
The comparison with \citet{Mannucci2005} data supports a value for the fraction of binary systems suitable 
to give rise to SNe~Ia of $0.03$, while disfavouring the value $0.1$ 
(at least for late-type galaxies and when the IMFs that we are adopting are considered).
Matching observational data of \citet{Mannucci2005} is a valuable result of our simulations: SN~rates 
are indeed a by-product of the chemical evolution network in our simulations, and reflect the past history 
of star formation and feedback of the galaxy (see also Table \ref{Success}). 
Also, the comparison between computed SN rates with observations can be used as a powerful tool to 
predict or at least validate the morphological type of the simulated galaxies.

\subsection{Metallicity profiles}
\label{MetallicityProfile}

\subsubsection{Stellar metallicity gradient}
\label{StellarMetallicityProfile}

\begin{figure}
\newcommand{\captionfonts}{\small}
\begin{minipage}{\linewidth}
\centering
\includegraphics[trim=0.3cm 0.cm 0.1cm 0.cm, clip, width=1.\textwidth]{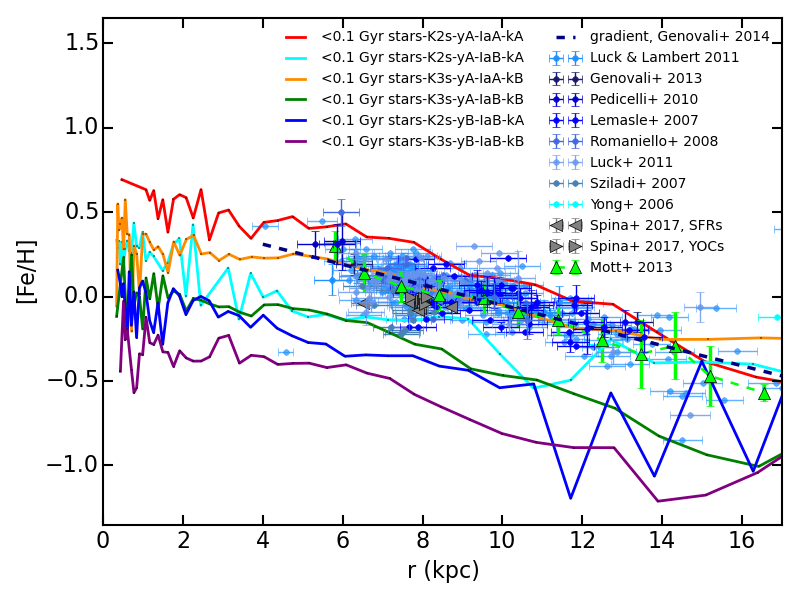} 
\end{minipage} 
\caption{Iron abundance radial profile for young stars in the simulated galaxies (colours as in the legend and previous 
	figures). We compare results with observations of Cepheids in the catalogue of \citet{Genovali2014yCat}, 
	that include data from \citet{Yong2006, Lemasle2007, Sziladi2007, Romaniello2008, Pedicelli2010, Luck2011_142_51L, Luck2011, Genovali2013}. Dashed blue line depicts the metallicity gradient for the whole 
	sample \citep{Genovali2014}. We also consider data for young open clusters (YOCs) and star 
	forming regions (SFRs) in MW from \citet{Spina2017}. Green symbols highlight observations of 
	\citet{Luck2011} according to the binning by \citet{Mott2013}. 
	}
\label{FeHstars} 
\end{figure}

\begin{figure}
\newcommand{\captionfonts}{\small}
\begin{minipage}{\linewidth}
\centering
\includegraphics[trim=0.3cm 0.cm 0.1cm 0.cm, clip, width=1.\textwidth]{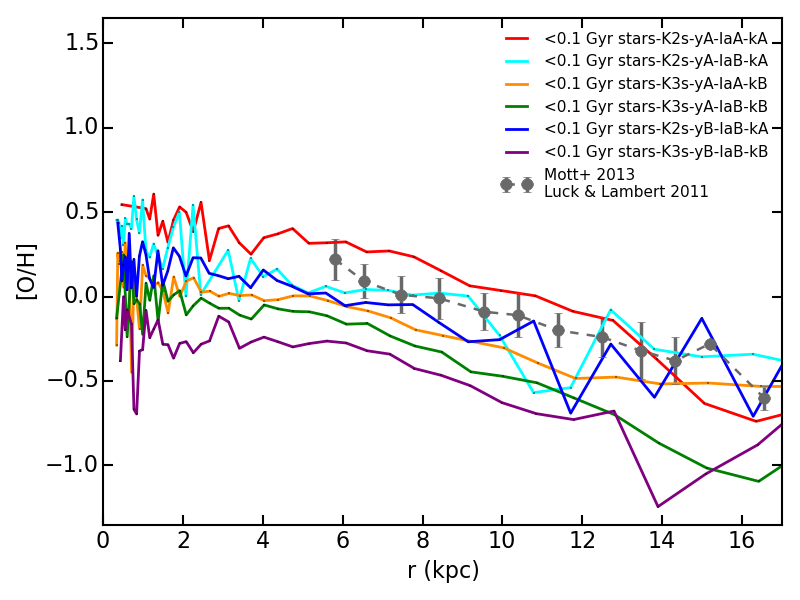} 
\end{minipage} 
\caption{Oxygen abundance radial profile for young stars in the set of galaxies. Data show the radial gradient for 
	oxygen from observations of Cepheids \citep{Luck2011}, according to the division in bins computed by 
	\citet{Mott2013}. Error bars show the standard deviation in each bin.}
\label{OHstars} 
\end{figure}

\begin{figure}
\newcommand{\captionfonts}{\small}
\begin{minipage}{\linewidth}
\centering
\includegraphics[trim=0.3cm 0.cm 0.1cm 0.cm, clip, width=1.\textwidth]{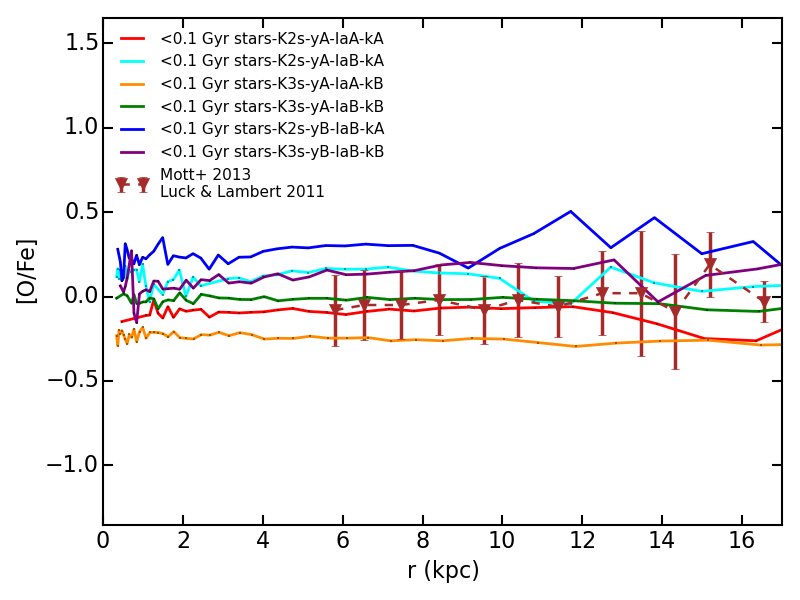} 
\end{minipage} 
\caption{Oxygen-over-iron abundance profile for young stars in simulated galaxies. Observations of Cepheids 
	from \citet{Luck2011}, according to the binning performed by \citet{Mott2013}.}
\label{OFestars} 
\end{figure}

In this section we analyse stellar metalicity gradients of simulated galaxies and compare predictions 
from simulations to observations, taking advantage of the wealth of accurate measurements available for our Galaxy. 
Figure \ref{FeHstars} shows the iron abundance radial profile for the set of simulated galaxies. Results are 
compared with observations of Cepheids in MW from different authors (further 
detailed in the legend and caption of the figure). Here, and in Figures \ref{OHstars} and \ref{OFestars}, we consider 
only those stars whose age is $< 100$ Myr in simulated galaxies, since Cepheids are stars younger 
than $\sim 100$ Myr \citep{Bono2005}. Ages of young open clusters and star forming regions in \citet{Spina2017} are 
younger than $\sim 100$ Myr, too (see Figure \ref{FeHstars}). When different from ours, solar reference abundances 
adopted in the observational samples have been properly rescaled.

Figure \ref{OHstars} shows the radial abundance gradient for oxygen. We compare predictions from our 
simulations with observations of Cepheids from \citet{Luck2011}, according to the division in bins as a function 
of the Galactocentric distance computed by \citet{Mott2013}. Error bars indicate the standard
deviation computed in each bin, that is $1$ kpc wide.

Figure \ref{OFestars} compares radial profiles for the abundance ratio [O/Fe] in simulations with 
observations of Cepheids  \citep{Luck2011, Mott2013}. Observations show that stars in the disc of MW are not 
$\alpha$-enhanced, on average, with respect to solar values. Data of \citet{Mott2013} have been 
shifted by the authors themselves (by the following amounts: $-0.17$ dex for [O/H], $-0.07$ for [Fe/H]) 
with respect to the original data of \citet{Luck2011}, due to the discrepancy of the dataset of \citet{Luck2011} 
with previous observations by the same authors (see below). 

Negative radial abundance gradients are recovered for all the galaxies, thus in agreement with observations 
(Figures~\ref{FeHstars} and \ref{OHstars}).
The slope of the profiles (for distances form the galaxy centre larger than $r > 4$ kpc) is in keeping with observations, 
this indicating the effectiveness of our sub-resolution model to properly describe the chemical
enrichment of the galaxy at different positions.

Focussing on the normalization of oxygen profiles, comparison with Cepheids supports a K2s IMF. 
The K2s IMF leads to a fair agreement with observations also when the normalization of iron profiles is considered: 
however, in this case, a value of $0.1$ for the fraction of binary systems with characteristics suitable to host 
SNe~Ia is the driver for matching observations. Indeed, when the value of $0.03$ is considered, the amount of 
iron produced is not enough to agree with data. 
The set B of stellar yields lowers the normalization of metallicity gradients, as a lower amount of oxygen and 
iron are synthesised (see Section~\ref{yields}). Iron profiles in galaxies simulated adopting this set of yields 
underestimate observations by $\gtrsim 0.5$ dex (Figure~\ref{FeHstars}).  
Also, there exists a degeneracy between the adopted IMF and the set of yields, the galaxy models 
K3s--yA--IaB--kB and K2s--yB--IaB--kA providing comparable results. 
Peaks at distances $r > 10$ kpc for galaxies K2s--yA--IaB--kA and 
K2s--yB--IaB--kA are due to the reduced amount of stars as the outermost regions of the stellar disc are reached. 

Figures \ref{FeHstars} and \ref{OHstars} suggest that young stars in these simulated galaxies are more metal poor 
than observed ones in MW, when both the iron and the overall metal content ($\sim$ oxygen) are considered. 
This is especially true for those models of galaxies that best fit the observations 
of gas oxygen abundance profiles 
in Figure \ref{OHgas} (see Section~\ref{GasMetallicityProfile}), 
and in particular for K3s--yA--IaB--kB (green). 
A possible reason for that stems from the star formation history of our galaxies. 
MW-like galaxies often assemble a considerable part of the stellar mass in the disc 
in the redshift range $2.5 \lesssim z \lesssim 1$ \citep{VanDokkum2013, snaith2014}. 
Our galaxies are instead characterised by a quiescent star formation in the redshift range 
$2 \lesssim z \lesssim 1$, as shown in Figure \ref{sfr}. 
Should a more continuous and sustained star formation at 
$2.5 \lesssim z \lesssim 1$ occur, a higher quantity of metals is produced and provided to the ISM, 
so that stars born after $z \sim 1$ would be much metal richer than they actually are. 

However, we would like to emphasise two relevant aspects that cannot be neglected and that could alleviate 
the tension between our predictions and observations. These are 
the uncertainties that affect metallicity determinations in observations and 
the peculiarity of Cepheids.

First, abundance estimate reliability is a delicate topic, as different sources of error enter final determinations 
\citep[see e.g.][]{Luck2011}. 
The quoted typical uncertainties for values of [Fe/H] in Cepheids catalogues considered in Figure \ref{FeHstars} 
are $0.1$ dex \citep{Romaniello2008} or $0.2$ dex \citep{Yong2006}.
Also, \citet{Luck2011} found a discrepancy of $0.07$~dex between the [Fe/H] derived from their sample 
and that computed for common objects in previous works by \citet{LuckKA2006, Luck2011_142_51L}. 
In these works they used equivalent analysis techniques, eventually tracing back to the changed surface gravity 
of stars the most plausible reason of the mismatch 
\citep[see the discussion in Section 4.1.1 of][for further details]{Luck2011}. 
The discrepancy was even larger ($\sim 0.17$~dex) when [O/H] is taken into account. 
Besides, variability of Cepheids can enter final estimates, that can suffer an uncertainty as high as $\sim 0.1$ dex due to 
the observed phase \citep{Luck2011}.

Also, different procedures adopted for metallicity calibrations yield remarkably different results. For instance, 
\citet{Kewley2008} quantified that different techniques aimed at determining abundances from 
metallicity-sensitive emission line ratios affect the normalization of profiles by up to $0.7$~dex. 
This, along with parameters in the sub-resolution physics that enter the modeling of the chemical enrichment process 
in simulations, usually lead different authors not to consider at all the normalization of the metallicity profiles or of the mass-metallicity relation. They rather focus on slopes only and, in case, on the evolution of the 
normalization of metallicity profiles \citep[e.g.][]{Torrey2017}. Therefore, differences in the normalization 
of profiles by up to few tens of dex should not be considered as indicative of a serious disagreement between 
predictions from simulations and observations. 

The second aspect concerns the possibility that Cepheids are 
stars with a metal content that is, on average, higher than what usually observed in nearby disc galaxies. 
This issue will be thoroughly addressed in Section~\ref{GasMetallicityProfile}.

\subsubsection{Gas metallicity gradient}
\label{GasMetallicityProfile}

\begin{figure}
\newcommand{\captionfonts}{\small}
\begin{minipage}{\linewidth}
\centering
\includegraphics[trim=0.3cm 0.cm 0.1cm 0.cm, clip, width=1.\textwidth]{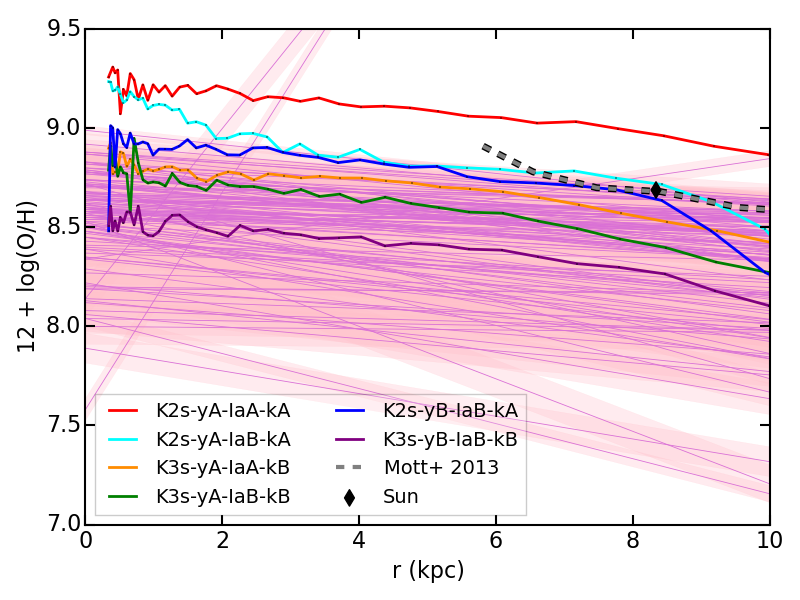} 
\end{minipage} 
\caption{Oxygen abundance radial profile for gas in the simulated galaxies (colours as in the legend and previous figures). 
	Light-blu profiles are observed metallicity gradients from the sample of $130$ nearby late-type galaxies of 
	\citet{Pilyugin2014}. The shaded envelope around each profile depicts the scatter of oxygen abundance around 
	the trend. We also show the metallicity gradient of MW (dashed curve) as deduced from 
	observations of young stars in our Galaxy, and the position of the Sun on the plane (see the text for details).}
\label{OHgas} 
\end{figure}

We continue our analysis by further investigating the metal content of gas in our galaxies.
Figure \ref{OHgas} shows the gas metal content of simulated galaxies. We analyse oxygen abundance gradients, 
oxygen being one of the most accurate tracers of the total metallicity.
We compare results from our set of simulations with the oxygen abundance radial profiles of the $130$ disc galaxies 
that make up the observational sample of \citet{Pilyugin2014}. The gas metal content of these late-type galaxies in the 
local universe has been inferred from spectra of H~II regions. 
Since we are not simulating our Galaxy, the comparison with properties of a set of disc galaxies 
is a key test.

All the simulated galaxies are characterised by negative gradients, in agreement with the majority of observed galaxies 
and with the scenario that outer regions of the galaxy discs are younger than inner ones 
(see also Section \ref{AgeRates}). Also, the slope of profiles of simulated galaxies is in keeping with the bulk of 
observations: this feature highlights the ability of our sub-resolution model to describe properly the local chemical 
enrichment process, as well as the circulation of metals driven by galactic outflows.

As for the amount of oxygen in the gas (i.e. the normalization of profiles), we find that the K2s IMF overproduces 
massive-star related metals. The oxygen profile of the galaxy K2s--yA--IaA--kA lies $\sim 0.3$ dex above 
the highest observed metallicity gradients, and the K2s--yB--IaB--kA model 
traces the upper edge of the observed metallicity range (see also below). 
On the other hand, we see that assuming the K3s IMF leads to a better agreement with observations 
by \citet{Pilyugin2014}. The K3s IMF limits the number of massive stars, thus resulting in a lower amount 
of massive-star related metals produced and provided to the ISM. 

The set B of stellar yields drifts towards a better agreement with observations
(compare, for instance, K2s--yB--IaB--kA and K3s--yB--IaB--kB models 
with K2s--yA--IaA--kA and K3s--yA--IaB--kB ones, respectively), 
as a lower amount of oxygen is produced. This result is more striking when contrasting the green and purple 
galaxies, that experience similar star formation histories. 

An interesting indication of our analysis is that a \citet{kroupa93} (K3s) IMF leads to a better agreement with metallicity 
profiles of late-type galaxies in the nearby universe shown in Figure~\ref{OHgas}. 
The comparison with the sample of \citet{Pilyugin2014} suggests that an IMF more top-light than 
\citet{Kroupa2001} (K2s) and \citet{ChabrierIMF2003} has to be preferred for local disc galaxies. 

However, the scenario becomes puzzling and the interpretation of results more challenging when we 
consider metallicity determinations for MW along with measurements in the sample of \citet{Pilyugin2014}. 
We overplot in Figure~\ref{OHgas} the oxygen abundance gradient of Cepheids in MW by 
\citet[][see the dashed grey line in Figure \ref{OHstars}]{Mott2013}.
Considering that the metal content of stars reflects directly that of gas out of which they formed, 
the metal content of young stars should be comparable to that of gas at redshift $z=0$, or should fall slightly below, 
at most. Also, we show where the Sun is located in Figure~\ref{OHgas}, considering 
$R_{\odot} = 8.33$ kpc as the distance of the Sun from the Galaxy centre \citep{Gillessen2009, Bovy2009}, and 
{log$_{10}~\varepsilon_{\rm O}~=~12~+~$log$_{10} \bigl(O/H\bigr)_{\odot}~=~8.69$} \citep{Asplund2009}. 
We see that the metal content of young stars in MW (that is a lower limit for the gas metallicity) outlines the 
extreme upper edge of the region where observations by \citet{Pilyugin2014} locate. 

This points out that either there are some issues with calibrations of metallicities in the considered observational 
samples, or Cepheids are stars with a metal content that is, on average, 
higher than what usually observed in nearby disc galaxies. 
If we deem the metallicity estimates of both samples as reliable, then we can conclude that the K2s IMF 
is more suitable for MW, while a K3s IMF should be preferred for other disc galaxies in the local universe. 

Also, it is striking that the Sun is characterised by an oxygen abundance comparable to that of Cepheids at 
the same distance from the Galaxy centre, although it is by far older than them. 
We envisage that the availability of more accurate observational determinations in the near future 
will alleviate the tension between discordant results and corroborate more robust conclusions.  
In Section \ref{otherResults} we further discuss how metals are retained and distributed in and around galaxies.

\subsection{Stellar $\alpha$-enhancement}
\label{StellarContent}

We further analyse the metal content of the stellar component of our galaxies focussing on the stellar 
enhancement in $\alpha$-elements with respect to the solar ratio.
Figure \ref{OsuFe-FesuH} shows [O/Fe] ratios as a function of [Fe/H] for the set of simulated galaxies. 
Each row refers to a galaxy model. Panels of the left column represent the distribution of all the star particles in 
the volume we are analysing ($r< R_{\rm gal}$, see Section \ref{featuresOfTheGalaxy}). Central and right 
columns show the distribution [O/Fe] ratios versus [Fe/H] for two subsamples of star particles, representative of 
stars located in the bulge and in the disc of galaxies, respectively. 

We sliced the whole sample of star particles according to the kinematic diagnostic indicator $J_{\rm z}/J_{\rm circ}$ 
(see Section \ref{featuresOfTheGalaxy}). We selected star particles with {$-0.1 < J_{\rm z}/J_{\rm circ} < 0.1$} as 
belonging to the bulge, and star particles with {$0.9 < J_{\rm z}/J_{\rm circ} < 1.1$} as located in the galaxy disc. 
Then, we also verified the position of selected particles, restricting the bulge subsample to star particles located at 
$r<3$ kpc from the galaxy centre, and the disc subsample to star particles whose distance from the galaxy centre 
is $3<r<13$ kpc and whose galactic latitude is $|z|<5$ kpc (as done in Figure \ref{StellarAgeProfiles}, for instance).
The number of star particles considered in second- and third-column panels of Figure \ref{OsuFe-FesuH} is not 
always the same, as galaxies have different stellar mass in the disc and the bulge (see Figure \ref{jcirc}). 

Particles selected according to these criteria sample specific components of the galaxy: their chemical features 
will be compared to observed stars that are located in the bulge or the disc of MW and that are characterised by 
peculiar evolutionary patterns.

\begin{figure*}
\newcommand{\captionfonts}{\small}
\includegraphics[trim=0.cm 0.1cm 7.cm 2.3cm, clip, angle=270, width=1.005\textwidth]{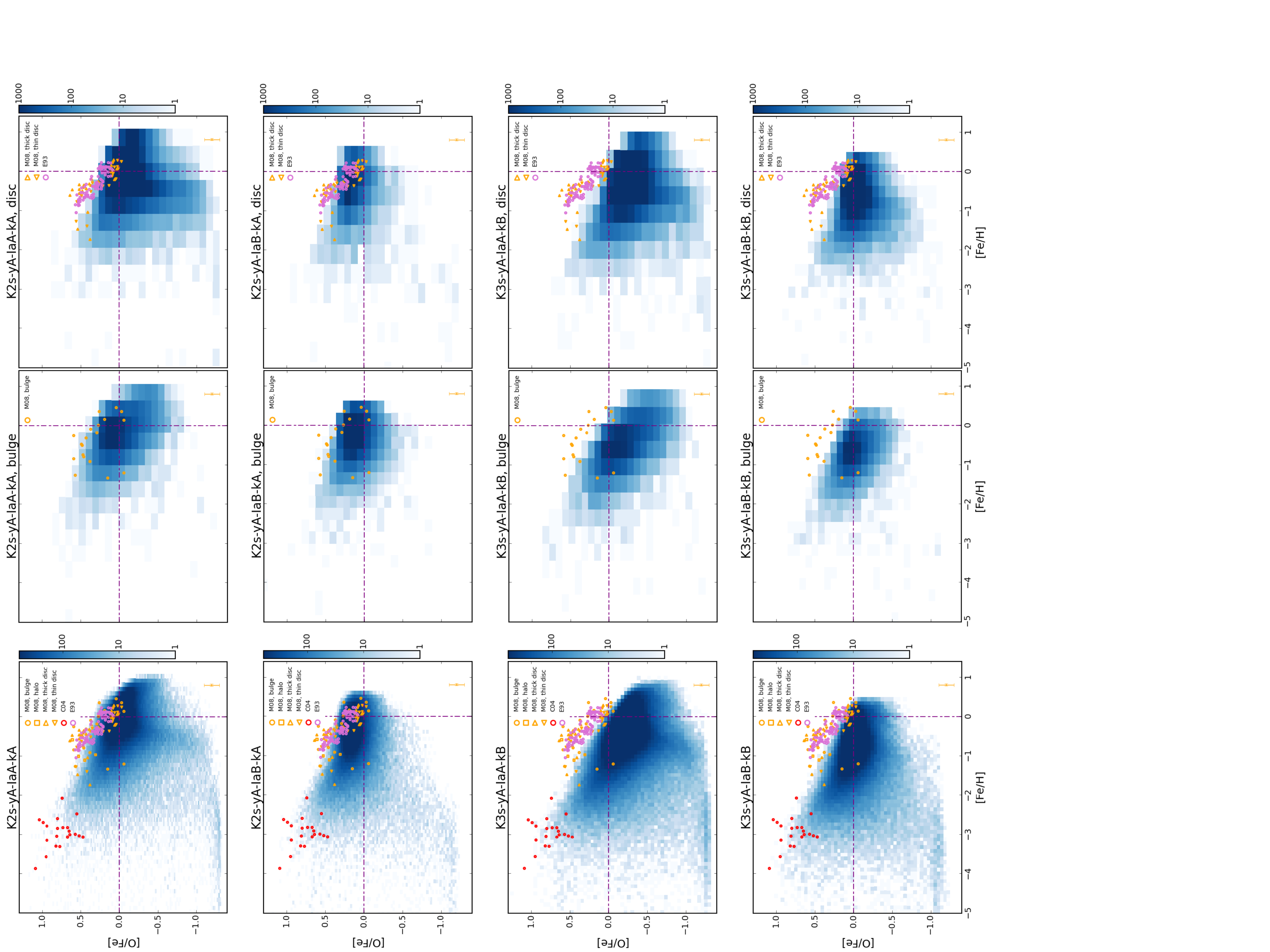}
\caption{[O/Fe] as a function of [Fe/H] for star particles in our simulations, at $z=0$. 
	Each row corresponds to a galaxy model. We show the [O/Fe] versus [Fe/H] for all the star particles within 
	$R_{\rm gal}$ (left-hand column), for star particles belonging to the bulge (centre), and for 
	star particles in the disc (right-hand column). Each colour bar encodes the number of star particles per pixel. 
	We contrast our results with observations of \citet[][M08]{Melendez2008}, of \citet[][C04]{Cayrel2004}, 
	and of disc stars by \citet[][E93]{Edvardsson1993}. 
	Dash-dotted violet lines highlight solar values. 
	A representative error bar for \citet[][M08]{Melendez2008} data is shown.} 
\label{OsuFe-FesuH} 
\end{figure*}
\addtocounter{figure}{+1}
\begin{figure*}
\ContinuedFloat
\includegraphics[trim=0.cm 0.1cm 21.5cm 2.3cm, clip, angle=270, width=1.005\textwidth]{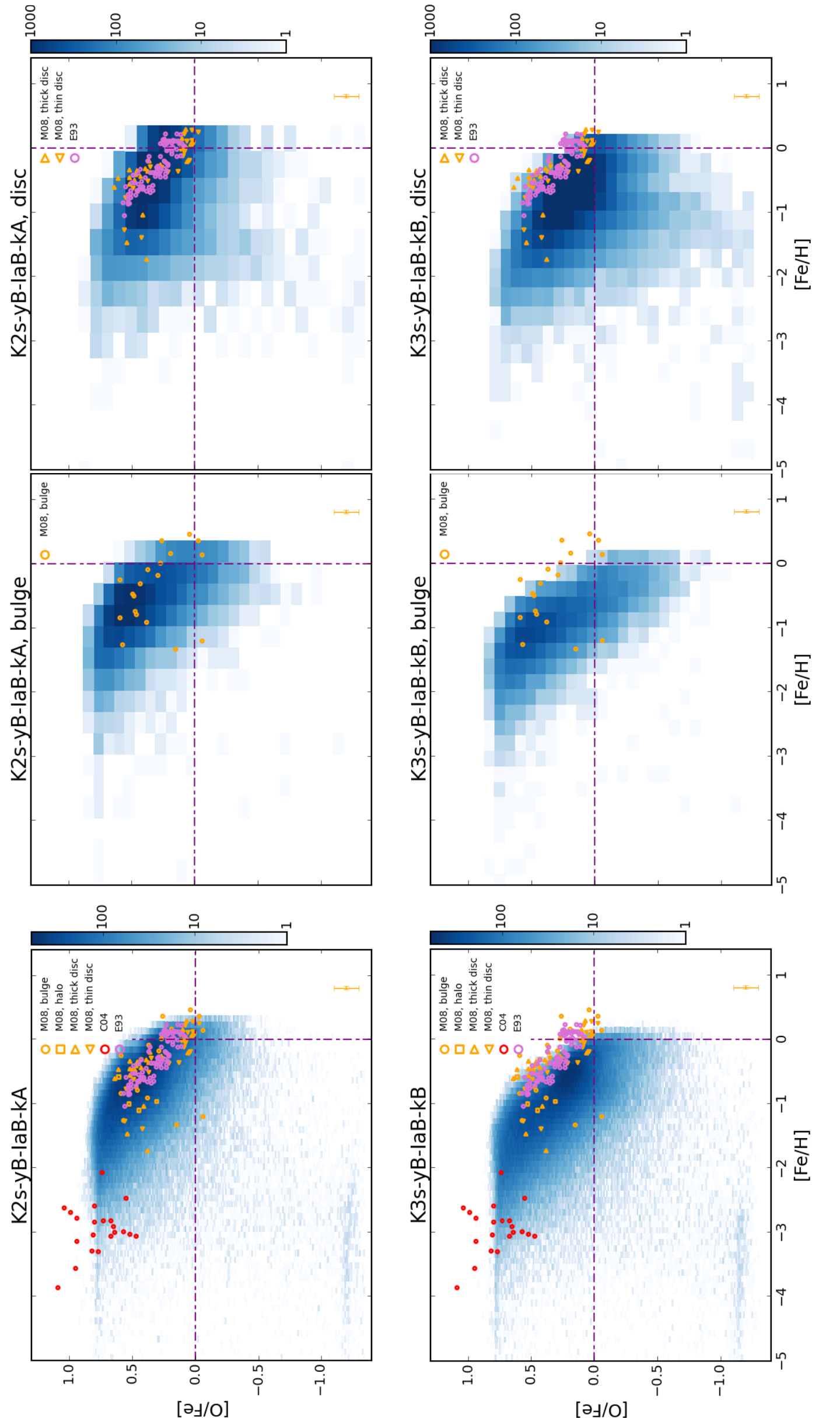}
\caption{continued} 
\end{figure*}

\begin{figure*}
\newcommand{\captionfonts}{\small}
\includegraphics[trim=0.cm 0.4cm 28.5cm 2.8cm, clip, angle=270, width=1.005\textwidth]{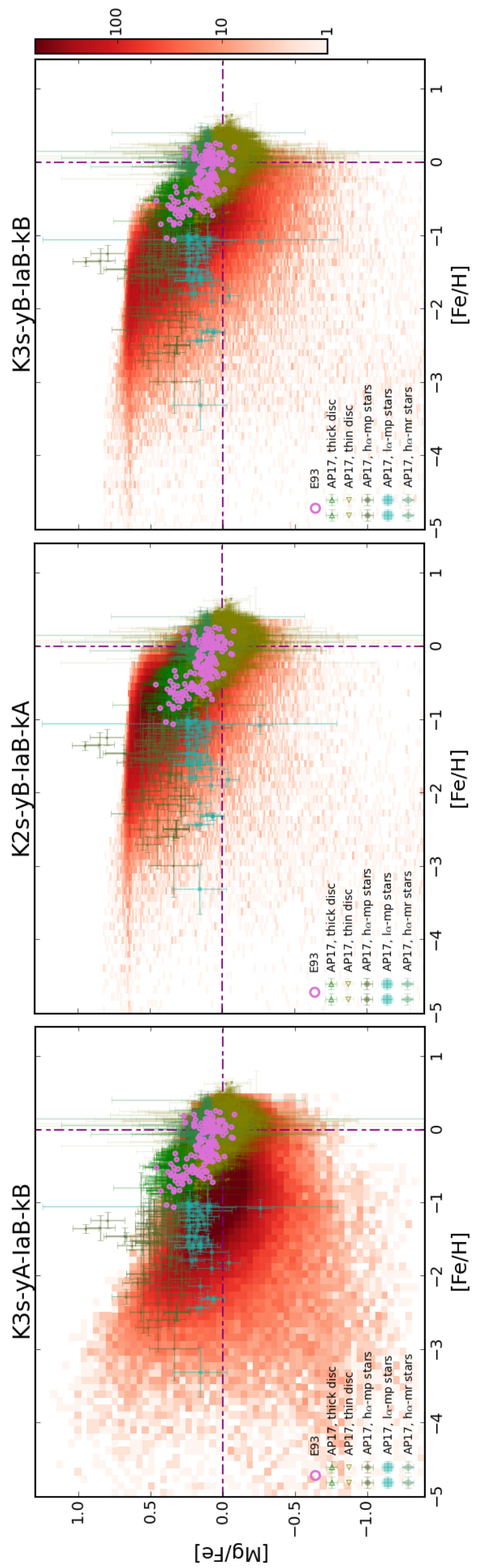}
\caption{[Mg/Fe] as a function of [Fe/H] for star particles in three simulations, at $z=0$. 
	Each panel corresponds to a galaxy model (green, blue and purple from left to right). 
	We show the [Mg/Fe] versus [Fe/H] for all the star particles 
	within $R_{\rm gal}$. Each colour bar encodes the number of star particles per pixel. 
	Results are compared with observations of the AMBRE project \citep[][AP17]{Mikolaitis2017}, 
	and of disc stars by \citet[][E93]{Edvardsson1993}. 
	Dash-dotted violet lines highlight solar values. 
	Data from the AMBRE project are split into thick and thin disc stars, high- and low-$\alpha$ metal-poor 
	population, and high-$\alpha$ metal-rich population. They are provided with error bars.} 
\label{MgsuFe-FesuH}  
\end{figure*}

Predictions from our simulated galaxies are compared with the following observational data set: 
{\sl {i)-}} \citet{Melendez2008}: they observed bulge, halo, thin and thick\footnote{Given the resolution of 
	our simulations, we do not attempt to resolve the structure of the simulated galaxy disc by distinguishing 
	between thick and thin disc.} disc giant stars; the quoted 
uncertainties of their data are $0.03$ dex in $Fe$ and $0.1$ dex in [O/Fe]. 
{\sl {ii)-}} \citet{Cayrel2004}: they investigated very metal-poor halo giant stars, and state that the error in their 
abundance estimates can be as large as $\sim 0.2$ dex.
{\sl {iii)-}} \citet{Edvardsson1993}: they observed dwarf stars in the disc of MW; quoted errors are $0.1$ dex for 
abundances relative to hydrogen, and uncertainties comparable or even smaller for abundances relative to iron. 
In second- and third-column panels we only consider observations for stars in the bulge or disc, respectively.

When different from ours, solar reference abundances adopted in the observational samples have been properly 
rescaled and homogenised.

The background two-dimensional histograms in Figure \ref{OsuFe-FesuH} consider the distribution of star particles 
in the plane [O/Fe] versus [Fe/H]. We compare results from simulations to observations of single stars under 
the assumption that the metal content of a star particle in simulations statistically reproduces the mean metallicity 
of the SSP that the star particle samples. We refer the reader to \citet{Valentini2018} for another possible method 
\citep[see also][]{Grand2018aurigaia}.

General trends that we observe by investigating results shown in Figure \ref{OsuFe-FesuH} and comparing 
galaxy models are the following.
Adopting the K3s IMF leads to an overall reduction of oxygen that is in contrast with observations of stars in MW. 
This is strikingly evident when contrasting results from K2s--yA--IaA--kA (first row) and from 
K3s--yA--IaA--kB (third row). Decreasing the number of systems suitable for hosting SNe~Ia does not 
shift the bulk of star particles towards values of [O/Fe] high enough to alleviate the tension.

As for the fraction of binary systems originating SNe~Ia, the comparison with observations in Figure \ref{OsuFe-FesuH} 
definitively rules out the value $0.1$, as it produces an excess of stars with low [O/Fe] and super-solar 
[Fe/H] that is not in keeping with observations.

A result that emerges from this analysis is that stellar yields affect the cooling process and are therefore 
a key driver of star formation, especially at high redshift. As a consequence, 
changing the adopted set of yields shapes directly the patterns of chemical evolution 
and stellar $\alpha$-enhancement. 
Two main features appear when the set B of stellar yields is adopted, regardless of the considered IMF: 
{\sl{(i)}} a lower amount of iron is produced (see Section \ref{yields}), and {\sl{(ii)}} the galaxy bulge is more 
$\alpha$-enhanced and forms over shorter timescales. 
By comparing second and fifth rows in Figure~\ref{OsuFe-FesuH}, we see that the bulk of stellar mass in the bulge 
of K2s--yB--IaB--kA (blue model) forms at earlier epochs (i.e. lower [Fe/H]) 
than K2s--yA--IaB--kA (cyan model), as also highlighted in the first panel of 
Figure~\ref{StellarAgeHisto}. Also, it has been highly enriched by the chemical 
feedback from massive stars (see below). A similar picture holds when K3s--yA--IaB--kB and 
K3s--yB--IaB--kB are considered, too. 

The set B of stellar yields allows to obtain the best agreement with observations, especially when the K2s IMF is 
assumed (fifth row of Figure~\ref{OsuFe-FesuH}). 
Results for the galaxy model K2s--yB--IaB--kA are in remarkable agreement with data: we find that 
stars in the bulge have [O/Fe] that span from typical values of halo stars ([O/Fe]$ \sim 0.4 - 0.6$) to 
slightly sub-solar values as the iron abundance approaches [Fe/H]$=0$. This trend is in keeping with findings 
by \citet{Zoccali2006, Melendez2008}. Also, the location of the break at [Fe/H]$ \sim -1 - -0.5$ is in 
agreement with observations. 
Bulge stars that shape this break highlight that the formation of the bulge occurred over a time longer than 
$\sim 1$ Gyr, so that SN~Ia contributed with iron to the enrichment of the ISM.
This feature is remarkably evident when the set B of stellar yields is adopted, while the transition to the regime 
where SN~Ia contribute to the enrichment is smoother for the set A.  
The bulk of stars in the disc with [Fe/H]$ \gtrsim -0.5$ are characterised, on average, by a lower 
$\alpha$-enhancement with respect to stars in the bulge. This is in agreement with the scenario according to which 
the formation of the disc in late-type galaxies occurs over longer timescales than the bulge component, that is 
conversely mainly affected by early feedback from massive stars \citep[see e.g.][]{Lecureur2007}. 
However, observed stars belonging to the bulge or to the disc of MW do not occupy distinct regions in the 
[O/Fe]--[Fe/H] plane, they rather overlap for almost all the considered values of [Fe/H] 
\citep[see e.g.][]{Melendez2008}. 
Adopting the set B of stellar yields also reduces the number of star particles characterised by sub-solar values of 
both [O/Fe] and [Fe/H], that do not have an observational counterpart.\footnote{The lack of an 
observational counterpart may be ascribed to the fact that these stars are rather rare. These results can be 
considered predictions from our simulations.}

\citet{FewCalura2014} explored the impact of the adopted IMF and binary fraction on the 
chemical evolution of a disc galaxy having a virial mass approximately half of 
that of our galaxies. Interestingly, they found that the slope of the [O/Fe]--[Fe/H] relation for stars 
characterised by [Fe/H]$ \gtrsim -1 $ is regulated by the value of the binary fraction, and that 
the more top-heavy \citet{Kroupa2001} IMF produces [O/Fe] larger by $\sim0.2$~dex with 
respect to the \citet{kroupa93} IMF, in keeping with our results. 

Figure \ref{MgsuFe-FesuH} further investigates the stellar $\alpha$-enhancement focussing on the [Mg/Fe] ratios 
as a function of [Fe/H], for three galaxies selected out of the set. We contrast our results with observations of the 
AMBRE project \citep[][]{deLaverny2013, Mikolaitis2017}, and of disc stars by \citet[][]{Edvardsson1993}. 
Trends observed and discussed for the stellar [O/Fe] versus [Fe/H] relation in Figure \ref{OsuFe-FesuH} are 
confirmed. Once again, we find a remarkable agreement with observations for the galaxy model K2s--yB--IaB--kA.


\begin{table*}
\centering
\begin{minipage}{\linewidth}
\caption{Relevant quantities of the simulated galaxies.
Column 1: simulation label.
Column 2: virial radius. Note that the galactic radius $R_{\rm gal} = 0.1\,R_{\rm vir}$. 
Column 3: ratio between gas mass and gas plus stellar mass within $R_{\rm vir}$. 
Column 4: mass of metals in gas over gas plus stellar mass within $R_{\rm vir}$.
Column 5: mass of metals in stars over gas plus stellar mass within $R_{\rm vir}$.
Column 6: ratio between gas mass and gas plus stellar mass within $R_{\rm gal}$. 
Column 7: mass of metals in gas over gas plus stellar mass within $R_{\rm gal}$.
Column 8: mass of metals in stars over gas plus stellar mass within $R_{\rm gal}$.}
\renewcommand\tabcolsep{4.45mm}
\begin{tabular}{@{}lccccccccc@{}}
\hline
Name   & $R_{\rm vir}$ (kpc)   & $\frac{M_{\rm gas}}{M_{\rm gas + stars}}$ & 
$\frac{M_{\rm Z,gas}}{M_{\rm gas + stars}}$  &   $\frac{M_{\rm Z,stars}}{M_{\rm gas + stars}}$  &  &  $\frac{M_{\rm gas}}{M_{\rm gas + stars}}$    &  $\frac{M_{\rm Z,gas}}{M_{\rm gas + stars}}$  &   $\frac{M_{\rm Z,stars}}{M_{\rm gas + stars}}$    \\
\hline
\hline
 & & \multicolumn{3}{l}{within $R_{\rm vir}$}  && \multicolumn{3}{l}{within $R_{\rm gal}$} \\
 \cline{3-5} \cline{7-9}
\hline
K2s--yA--IaA--kA & $238.7$   &    $0.77$      &    $2.8 \cdot 10^{-3}$     &   $5.6 \cdot 10^{-3}$  && $0.37$ & $7.6 \cdot 10^{-3}$     &   $1.7 \cdot 10^{-2}$\\  
\hline
K2s--yA--IaB--kA & $236.8$   &  $0.84$      &    $1.4 \cdot 10^{-3}$     &     $2.6 \cdot 10^{-3}$  && $0.39$ & $4.9 \cdot 10^{-3}$     &   $1.1 \cdot 10^{-2}$ \\  
\hline
K3s--yA--IaA--kB & $241.5$   &  $0.81$      &    $1.5 \cdot 10^{-3}$     &     $1.9 \cdot 10^{-3}$  && $0.42$ & $3.9 \cdot 10^{-3}$     &   $6.0 \cdot 10^{-3}$ \\  
\hline
K3s--yA--IaB--kB & $240.3$   &  $0.85$      &    $8.3 \cdot 10^{-4}$     &     $9.9 \cdot 10^{-4}$  && $0.47$ & $2.7 \cdot 10^{-3}$     &   $3.9 \cdot 10^{-3}$ \\  
\hline
K2s--yB--IaB--kA & $236.9$   &  $0.84$      &    $1.1 \cdot 10^{-3}$     &     $1.9 \cdot 10^{-3}$  && $0.33$ & $3.7 \cdot 10^{-3}$     &   $8.9 \cdot 10^{-3}$ \\  
\hline
K3s--yB--IaB--kB & $239.6$   &  $0.87$      &    $5.8 \cdot 10^{-4}$     &    $6.9 \cdot 10^{-4}$  && $0.47$ & $2.0 \cdot 10^{-3}$     &   $3.2 \cdot 10^{-3}$\\  
\hline
\hline
\end{tabular}
\label{Numbers}
\end{minipage}
\end{table*}


Results in this section show a better agreement with observations of stars in our Galaxy when the K2s IMF is 
assumed. On the other hand, the comparison of gas metallicity profiles with observations of a sample of disc 
galaxies in the local universe (Section \ref{GasMetallicityProfile}) strongly supports the K3s IMF. Therefore, 
in agreement with our previous findings in Section \ref{StellarMetallicityProfile}, we predict that MW stars exhibit a 
higher metal content with respect to other nearby late-type galaxies. Interestingly, stars of MW are found to be 
more $\alpha$-enhanced than stars in nearby dwarf galaxies, and not representative of stellar populations in dwarfs 
\citep[][and references therein]{Venn2004}, even if different formation scenarios can enter this result.

We stress once again that our goal is not to reproduce observations of the MW and that we simulate a disc galaxy 
that should not be considered as a model for the MW. 
We are rather interested in investigating whether an accurate and controlled model of chemical evolution 
as the one we adopt in our simulations is able to yield chemical evolutionary patterns 
that are associated to specific components and timescales by observations.

\begin{figure}
\newcommand{\captionfonts}{\small}
\begin{minipage}{\linewidth}
\centering
\includegraphics[trim=0.3cm 0.cm 0.1cm 0.cm, clip, width=1.\textwidth]{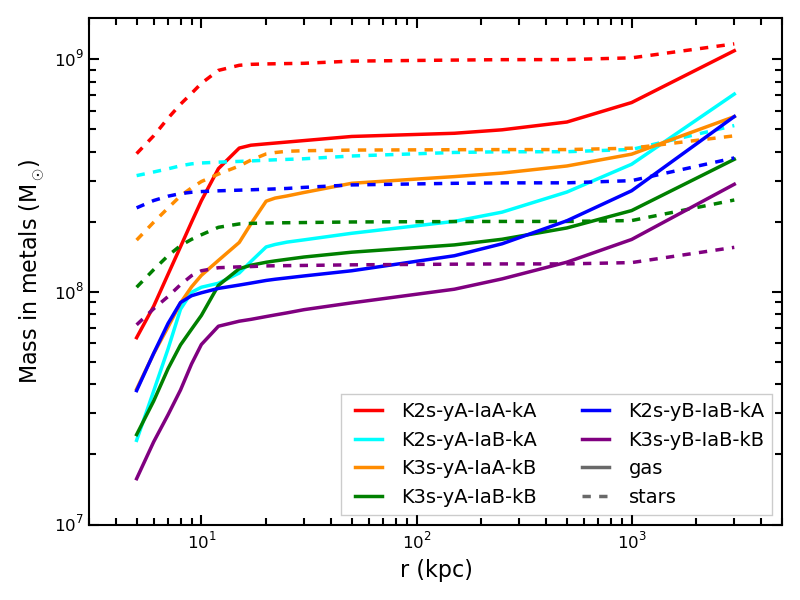} 
\end{minipage} 
\caption{Cumulative radial distribution of metals in gas (solid lines) and stars (dashed lines) in our 
simulated galaxies. Each curve shows the mass of metals in gas and stars enclosed within a given distance 
from the galaxy centre. }
\label{whereMetals} 
\end{figure}

\subsection{Metals in and around galaxies}
\label{otherResults}

In this section we focus on the spatial distribution of metals in and around galaxies as the result 
of chemical feedback from outflows. 
Comparing predictions from our simulations with observations of the CGM is beyond the scope of this paper, 
as this kind of analysis would require an accurate modelling of the ionization state of the gas. 
Here we want to show that the high galactic metal content reached when the K2s IMF is assumed 
(see Figure~\ref{OHgas}) is not due to the ineffectiveness of galactic outflows in removing metals from 
sites of star formation, within the innermost regions of galaxies. 
To pursue that, we analyse how the mass budget of metals is distributed at increasing distance from the galaxy centre.

Figure \ref{whereMetals} shows the cumulative distribution of metals in gas and stars of simulated galaxies 
as a function of the distance from the galaxy centre. 
We consider $3$~Mpc as the outermost distance within which to compute the mass of metals as this is the radius of 
the largest sphere that can be contained in the volume with high resolution particles (see Section \ref{ICs}).

Metals have been produced by stars in the main galaxy sitting at the centre of the simulated halo and by stars 
in satellites and few small field galaxies. The latter stars are responsible for the mild increase of the mass of metals 
in stars at distances larger than the galactic radius ($R_{\rm gal} \simeq 24$~kpc) of the main galaxies. Radial 
profiles of mass of metals in stars in Figure \ref{whereMetals} smoothly increase up to reach the stellar extent of 
galaxies, and then marginally rise when other stars not associated to the main galaxies are met in the volume. 

Radial distribution of metals in gas show how galactic outflows are effective in driving metals outwards from 
star-forming regions in galaxies and in promoting the metal enrichment of the ISM and CGM. The mass of metals in 
gas rapidly increases in the ISM of the galaxy up to distances $r \simeq 20$~kpc; from there outwards the mass of 
metals in gas increases by up to a factor $\sim 2 - 3$. 
Outside $R_{\rm gal}$ metals are retained by the diffuse CGM. 
We find that the mass of metals in gas within a distance of $20$~kpc from the galaxy centre is {$18 \pm 4 \%$} 
of the total amount of metals produced (within $3$ Mpc\footnote{We are therefore quoting upper limits.}) 
and {$31 \pm 9 \%$} of the metals produced which belong to the gaseous phase.
Metals in gas and stars within $20$~kpc account for at least $\sim 40 \%$ of the metals produced within $3$~Mpc. 
The mass of metals in gas and stars within $150$~kpc from the galaxy centre constitutes $57 \pm 9 \%$ of the metals 
produced. This result is in remarkable agreement with \citet{Peeples2014}, which investigated a sample 
of nearby star-forming galaxies with stellar mass in the range $10^9 - 10^{11.5}$~M$_{\odot}$ and found 
that galaxies retain $\sim 50 \%$ of produced metals in their ISM/CGM and stars out to $150$ kpc, 
with no significant dependence on the galaxy mass.

\begin{figure*}
\newcommand{\captionfonts}{\small}
\begin{minipage}{\linewidth}
\centering
\includegraphics[trim=0.2cm 0.2cm 0.2cm 0.2cm, clip, width=.49\textwidth]{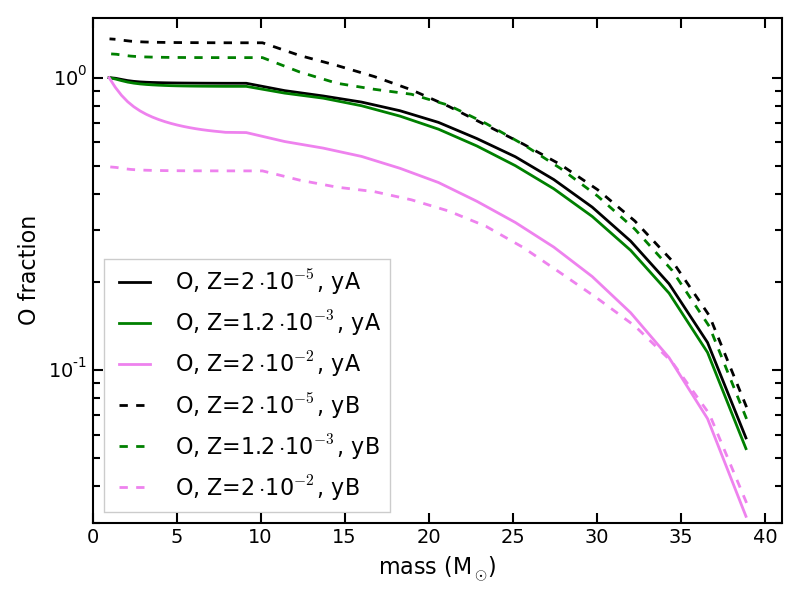}
\includegraphics[trim=0.2cm 0.2cm 0.2cm 0.2cm, clip, width=.49\textwidth]{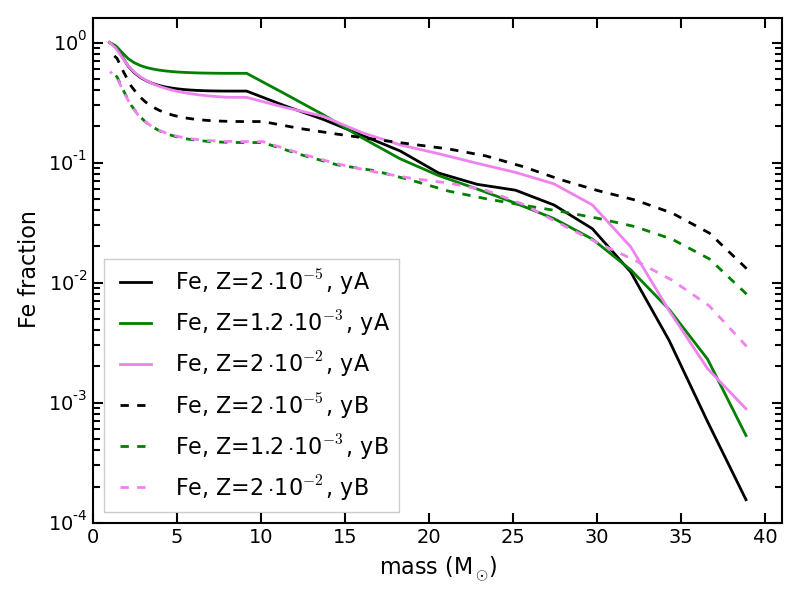}
\end{minipage} 
\caption{Cumulative mass in oxygen (left-hand panel) and iron (right-hand panel) that is (newly) produced by an 
	SSP as a function of the mass of stars in the SSP. Solid and dashed curves refer to the sets A and B 
	of stellar yields, respectively. Curve of each considered metallicity are normalised to the maximum of the curve 
	of the set A. We consider the K3s IMF. 
	We analyse three possible values for the metallicity $Z$: $2 \cdot 10^{-5}$ (black), 
	$1.2 \cdot 10^{-3}$ (green), and $2 \cdot 10^{-2}$ (pink).}
\label{FeOyields_setAB} 
\end{figure*}

By analysing the mass distribution of metals in gas and stars of our galaxies, we find the following results. The mass 
of metals locked in stars within $10$ kpc from the galaxy centre is at least twice as high as that in gas (mass ratios 
range between $2.22$ and $3.42$). The mean star-over-gas metal mass ratio $20$~kpc far from the galaxy centre 
is $1.96 \pm 0.40$, while at $50$ kpc it is $1.80 \pm 0.41$. A minimum distance of $\sim 500$~kpc 
(i.e. beyond the galaxy virial radius) has to be reached in order to have a comparable mass of metals 
retained by stars in the galaxy and present in the ISM and CGM of the galaxy. 
Table \ref{Numbers} summarises relevant quantities of the simulated galaxies. 

Also, the radial distribution of metals in gas shown in Figure~\ref{whereMetals} sheds further light on the metal 
content of gas within the galaxies shown in Figure \ref{OHgas}. Galaxy models assuming the K2s IMF are 
characterised by an oxygen abundance of gas that is higher than commonly observed in disc galaxies in the local 
universe. This feature cannot be ascribed to the inability of galactic outflows to drive metals outward from 
sites of star formation within the galaxy. Conversely, outflows in galaxies adopting the K2s IMF are as effective as 
those in models with the K3s IMF, but an overproduction of metals occurs when the K2s IMF is assumed 
(see Section \ref{GasMetallicityProfile}). Thus, our prediction that a \citet{kroupa93} IMF seems to be preferred 
when dealing with nearby late-type galaxies is further corroborated.

\subsection{Stellar yields}
\label{yields}

In this section we quantify the impact of the adopted set of stellar yields. We analyse the mass of oxygen and iron 
that is produced by an SSP according to the two different yield tables, once an IMF is assumed. 
We consider the K3s IMF in this section.

Figure \ref{FeOyields_setAB} shows the normalised cumulative mass in oxygen (left-hand panel) 
and iron (right-hand panel) that is produced by an SSP as a function of the mass of stars in the SSP. 
We consider newly produced elements for three values of the SSP metallicity $Z$: 
$2 \cdot 10^{-5}$, $1.2 \cdot 10^{-3}$, and $2 \cdot 10^{-2}$. 
The mass range spans between the minimum mass of the K3s IMF, 
that is $0.1$~M$_{\odot}$ and $40$~M$_{\odot}$ (see Section \ref{suite}). 
In each panel, solid curves identify the set A of stellar yields, dashed curves pinpoint the set B. 
As for the normalization, each curve is normalised to the maximum of the curve of that metallicity 
referred to the set A. In this way, it is possible to appreciate which set of yields predict a 
larger amount of the considered element, at a given abundance. 

Considering a star of a given mass, these figures describe the cumulative mass of an element 
(oxygen or iron) that is produced by stars with mass equal or larger than the considered one. 
For instance, stars more massive than $15$~M$_{\odot}$ and with a roughly solar metallicity $Z=2 \cdot 10^{-2}$ 
have produced $\sim 55 \%$ of the oxygen that has been synthesised, when the set A is considered. 

The effect of changing yields on the metals produced by stars of different mass and initial metallicity is 
evident: different amounts of heavy elements are produced and this has significant consequences on the 
gas cooling and subsequent star formation. The lower the metal budget, the lower the cooling rate and 
the star formation rate, as discussed in Section \ref{featuresOfTheGalaxy}\footnote{
We postpone the analysis of the time evolution of the cooling rate for models adopting different sets of 
stellar yields to a future study. 
}.
When the set B of stellar yields is adopted, a lower amount of iron is produced for all the considered metallicities. 
The mass of synthesised oxygen is lower than in the set A for SSPs with solar abundance, and slightly higher 
for lower values of $Z$. However, we note that the bulk of young stars are characterised, on average, by a 
roughly solar metallicity (see Figure~\ref{OHstars}): as a consequence, also a lower amount of oxygen 
is produced when the set B is adopted, at least at low redshift. 
Total masses of oxygen and iron produced by SSPs with increasing metallicity increase when the set A of yields 
is adopted. When considering the set B of stellar yields, the total mass of iron keeps almost unaltered as a function 
of the metallicity $Z$, while the cumulative mass of oxygen slightly decreases as the metallicity increases.

\section{Discussion and Conclusions}
\label{sec:conclusions}

In this work we carried out a set of cosmological hydrodynamical simulations of disc galaxies, with zoomed-in 
initial conditions leading to the formation of a halo of mass $M_{\rm halo, \, DM} \simeq 2 \cdot 10^{12}$~M$_{\odot}$ 
at redshift $z=0$. 
These simulations have been designed to investigate the distribution of metals in galaxies and to quantify 
the effect of {\sl{(i)}} the assumed IMF, {\sl{(ii)}} the adopted stellar yields, and {\sl{(iii)}} the impact of binary systems 
originating SNe~Ia on the process of chemical enrichment. 
We considered either a \citet[][K3s]{kroupa93} or a \citet[][K2S]{Kroupa2001} IMF, either a value $0.1$ or $0.03$ for the 
fraction of binary systems suitable to give rise to SNe~Ia, and two sets of stellar yields.

The most relevant results of our study can be summarised as follows.
\begin{itemize}
\item We examined stellar ages and naturally predict that star formation in the disc always 
takes place later than in the bulge, and the disc formation proceeds inside-out.

\item We evaluated the evolution of SN~rates in the set of simulated galaxies and compared them to observations 
of \citet{Mannucci2005}. We find that SN~II rates of our galaxies agree with observations from late-type Sbc/d galaxies. 
Also, SN~Ia rates of simulations adopting $0.03$ as the fraction of binary systems that host SNe~Ia are in 
keeping with observations of Sbc/d galaxies. 

\item We investigated stellar and gas metallicity gradients in simulations: negative radial abundance profiles 
are recovered, thus in agreement with observations. Slopes of profiles agree with observations, pointing to the 
effectiveness of our model to properly describe the chemical enrichment of the galaxy at different positions.

\item We compared gas metallicity profiles from our set of simulations with the oxygen abundance radial profiles 
of $130$ nearby disc galaxies \citep{Pilyugin2014}. We found that a \citet{kroupa93} IMF leads to a lower 
amount of metals produced by massive stars and supplied to the ISM, resulting in a 
better agreement with observations. Therefore, we can argue that an IMF more top-light than 
\citet{Kroupa2001} and \citet{ChabrierIMF2003} has to be preferred for local disc galaxies. 

\item We investigated the stellar $\alpha$-enhancement of star particles in our simulations, and considered also 
chemical patterns associated to disc and bulge, separately. The best agreement with observations of 
MW stars is retrieved when the \citet{Kroupa2001} IMF is assumed. 

\item We quantified the effect of chemical feedback from outflows, investigating the mass of metals that is 
distributed at increasing distance from the galaxy centre. 
We verified that galactic outflows in our simulations are effective in driving metals out from the sites
of star formation: if a high amount of metals in the ISM is present, this cannot be ascribed to the inability of 
outflows to move enriched gas outwards. 
We found that the mass of metals in gas within a distance of $20$~kpc from the galaxy centre is $\sim 20 \%$ 
of the metals produced, 
while about as many metals are associated to stars within the same distance. 
The mass of metals in gas and stars within $150$ kpc from the galaxy centre increases to the $\sim 60 \%$ 
of the synthesised metals. Therefore, an amount of metals as large as $\sim 40 \%$ has been driven 
beyond $150$ kpc by galactic outflows, that have promoted the circulation of metals in the CGM at high redshift, 
during the galaxy formation process. 
\end{itemize}

In the light of these results, we can draw the following conclusions on the different ingredients entering in 
the adopted model of stellar evolution. 
\begin{itemize}
\item Assuming either a \citet{kroupa93} IMF or a \citet{Kroupa2001} IMF results in different star formation 
histories, stellar feedback and scenarios of chemical evolution. 

\item If observational metallicity estimates are not affected by significant uncertainties,
the \citet[][]{Kroupa2001} IMF is more suitable for our Galaxy, 
while a \citet[][]{kroupa93} IMF (more top-light) should be preferred for other disc galaxies in the local universe.
We predict that MW stars exhibit a higher metal content with respect to other nearby late-type galaxies. 

\item Our analysis supports a value for the fraction of binary systems suitable to give rise to SNe~Ia of $0.03$.  
This rules out the commonly adopted value $0.1$, at least for late-type galaxies and 
when the IMFs that we are adopting are considered. Evidences that corroborate this finding come from 
the comparison of predicted SN rates and stellar $\alpha$-enhancement with observations. 

\item Adopted stellar yields are a key component of the chemical model, as they control cooling and drive 
star formation. Different yields lead to different chemical features. The pattern of chemical enrichment for stars 
located in the galaxy bulge shows that the bulge component is the most sensitive to stellar yields. 
\end{itemize}

As already pointed out, the availability of further accurate data from ongoing surveys
could alleviate the tension between discrepant results and lead to firmer conclusions.
As a concluding remark, we stress that an emerging piece of evidence of our work is that 
it is not trivial to simulate galaxies with a dominant disc component and an almost quiescent low-redshift 
star formation rate that simultaneously match all the available observational constraints 
on the pattern of metal enrichment of both the gas and stellar components. 
Also, results are far from being predictable when a single ingredient entering in the model for chemical evolution 
is varied, as its effect has a complex interplay with processes 
regulating the star formation history and shaping the morphology of simulated galaxies. 
Our results highlight that it is challenging to reproduce at the same time observations of metal abundance 
in gas and stars. Observational uncertainties and issues with the calibration of metallicity estimates 
could definitely enter and complicate the comparison with simulation predictions. 
However, our analysis also highlights that the numerical modelling of the share 
of metals between stars and the surrounding gas is of paramount importance, and should be further 
investigated. To pursue that, higher resolution simulations are mandatory, 
as well as a proper treatement of the diffusion of metals associated to turbulent gas motions 
\citep[see e.g.][]{Pilkington2012, Williamson2016}. 

The detailed and accurate investigation of the chemical evolution of a single galaxy that we have presented 
in this work is the starting point of a systematic analysis that we plan to accomplish in future works. 
A natural and desirable extension of this study is indeed the simulation of cosmological boxes, where 
statistical properties of a population of galaxies can be examined. 

As a word of caution, we note that the simulations presented in this work do not include the effect of the AGN feedback. 
Albeit the AGN feedback is not expected to play a major role in the overall evolution of late-type galaxies nor 
to influence the history of chemical enrichment of the galaxy significantly, it can however play a supporting role 
in fostering outflows and promoting the circulation of metals over galactic scales. 
We postpone the investigation of the effect of the AGN feedback to a forthcoming paper.

\section*{Acknowledgments}
We thank the anonymous referee for the careful and constructive report that helped 
improving the presentation of results. 
We are greatly indebted to Volker Springel for giving us access to the
developer version of the GADGET3 code. 
We are grateful to Gabriella De Lucia for fruitful discussions and useful comments on the manuscript. 
We thank Donatella Romano for kindly providing us with stellar yields, Emanuele Spitoni for sharing 
observational data of radial abundance gradients of Cepheids, 
and Francesco Calura for useful discussions. 
Simulations were carried out using ULISSE at SISSA and Marconi at CINECA (Italy). CPU time has been 
assigned through the project Sis18\_bressan under Convenzione SISSA,
through the project INA17\_C1A00, 
and through Italian Super-Computing Resource Allocation (ISCRA) proposals and an agreement 
with the University of Trieste. 
The post-processing has been performed using the PICO HPC cluster at CINECA through our expression of interest. 
LT has been funded by EU ExaNest {\sl{FET-HPC}} project No 671553.




\bibliographystyle{mnras} 
\bibliography{cool_ref}





\bsp	
\label{lastpage}
\end{document}